\documentclass{article}
\usepackage[utf8]{inputenc}
\usepackage{amsmath}
\usepackage{graphicx}
\usepackage{hyperref}
\usepackage[left=2cm,right=2cm,
    top=2cm,bottom=3cm,bindingoffset=0cm]{geometry}

\title{Improving the Efficiency of Oncological Diagnosis of the Breast Based on the Combined Use of Simulation Modeling and Artificial Intelligence Algorithms \footnote{Published: Khoperskov A. V., Polyakov M. V. Improving the Efficiency of Oncological Diagnosis of the Breast Based on the Combined Use of Simulation Modeling and Artificial Intelligence Algorithms // Algorithms. 2022, Vol. 15, no. 8, 292. DOI: \url{https://doi.org/10.3390/a15080292}}}
\author{Alexander V. Khoperskov and Maxim V. Polyakov}

\date{\textit{Volgograd State University, Universitetsky pr., 100, Volgograd 400062, Russia}}

\begin{document}

\maketitle

\begin{abstract}
The work includes a brief overview of the applications of the powerful and easy-to-perform method of Microwave Radiometry (MWR) for the diagnosis of various diseases. The main goal of this paper is to develop a method for diagnosing breast oncology based on machine learning algorithms using thermometric data, both real medical measurements and simulation results of MWR examinations.
The dataset includes distributions of deep and skin temperatures calculated in numerical models of the dynamics of thermal and radiation fields inside a biological tissue. 
The constructed combined dataset allows us to explore the limits of applicability of the MWR method for detecting weak tumors. We use convolutional neural networks and classic machine learning algorithms (k-nearest neighbors, naive Bayes classifier, support vector machine) to classify data.
The construction of Kohonen self-organizing maps to explore the structure of our combined dataset demonstrated differences between the temperatures of patients with positive and negative diagnoses.
Our analysis shows that the MWR can detect tumors with a radius of up to 0.5 cm if they are at the stage of rapid growth, when the tumor volume doubling occurs in approximately 100 days or less. The use of convolutional neural networks for MWR provides both high sensitivity ($sens=0.86$) and specificity ($spec=0.82$), which is an advantage over other methods for diagnosing breast cancer. New modified scheme for medical measurements of IR temperature and brightness temperature is proposed for a larger number of points in the breast compared to the classical scheme. This approach can increase the effectiveness and sensitivity of diagnostics by several percent. 

\vspace{2mm}
\end{abstract}

\section{Introduction}
   
Temperature distributions inside biological tissues/organs and on the human body surface can be a source of important information about the functional state of the organism and negative processes associated with the development of the disease at the earliest \mbox{stages \cite{Moloney-etal-2022Wavelia-System, Levshinskii-etal-2019politeh, Stauffer-etal-2014brain-temperature, Osmonov-etal-2021coronavirus, Fasoula-etal-2021Wavelia-System, Moloney-2020review-Breast-Cancer-Detection}.} 
Measurement of passive radiation in the microwave range is a simple, non-invasive, cheap and fairly accurate method for determining temperature distributions inside biological tissues \cite{Levshinskii-etal-2019politeh, Robert-etal-1979Millimeter-wave-thermography, Levshinskii-etal-2022venous}.
This approach is called Microwave Radiometry (MWR).
The other terms are used in the literature to refer to this method, for example, Microwave Thermography, Microwave Thermometry, Microwave Radiothermometry, Microwave Thermo\-graphic and such studies of internal temperature began more than 40 years ago on examples of the breast \cite{Barrett-Myers-1975first-work-breast, Barrett-etal-1977breast-cancer-microwave}, knee joint \cite{Edrich-Smyth-1978Arthritis-thermography}, the head and the neck \cite{Robert-etal-1979Millimeter-wave-thermography}.

Significant advances in the use of microwave radiometry are associated with the diagnosis of breast cancer \cite{Moloney-etal-2022Wavelia-System, Levshinskii-etal-2019politeh, Fasoula-etal-2021Wavelia-System, Moloney-2020review-Breast-Cancer-Detection, Barrett-etal-1977breast-cancer-microwave, Polyakov-etal-2021ourVest, Germashev-2022Cyber-Physical-System, Ekici-Jawzal-2020Breast-cancer-neural-networks}.
A positive outcome of breast cancer treatment is determined by the earliest possible diagnosis, covering a significant part of the population \cite{Siegel-etal-2021Cancer-Statistics-2021}.
Since breast oncology dominates in women's mortality from cancer, the organization of mass early diagnosis and screening is a priority, which requires reliable algorithms for processing thermometric data.
Various aspects of the application of Microwave radiometry methods are actively discussed in a number of reviews \cite{Moloney-2020review-Breast-Cancer-Detection, Losev-Svetlov-2022, Paulides-etal-2020-review-hyperthermia, Schiavon-etal-2021Review-Joint-Diseases, Hassan-El-Shenawee-2014, Sowmya-etal-2022Microwave-Sterilisation}.  

Machine learning and the application of the full range of artificial intelligence algorithms are a powerful tool for processing medical measurement data \cite{Osmonov-etal-2021coronavirus, Ekici-Jawzal-2020Breast-cancer-neural-networks, Liang-etal-2022magnetic-resonance-imaging, Losev-etal-2022Machine-Learning, Ragab-etal-2021breast-cancer-Multi-DCNNs-mammogram}.
Convolutional neural networks can automatically detect signs of breast cancer in the thermal images with up to 98.95 percent accuracy \cite{Ekici-Jawzal-2020Breast-cancer-neural-networks}.
Microwave imaging technology makes it possible to monitor the progress of a disease in quasi-real time, such as brain stroke \cite{Ullah-Arslan-2021Microwave-Imaging-Algorithms, Scapaticci-etal-2012microwave-imaging-brain}.
Microwave radiometry provides clinical monitoring of brain temperature distribution during surgical procedures \cite{Stauffer-etal-2014brain-temperature}.

A particular task is the visualization of temperature fields during thermal therapy based on microwave ablation at frequencies of 300 MHz --- 300 GHz with local heating up to 60\,$^\circ$C and even higher in real time mode. This method allows destroying malignant tumors and their metastases. The procedure uses an antenna needle that is inserted into the tumor node and emits microwaves, resulting in intense localized heating \cite{Huang-etal-2014microwave-hyperthermia-antenna-needles}. This approach makes it possible to successfully treat a tumor up to 0.5~cm \cite{Witte-Tamimi-2022thermo-therapy, Moon-Brace-2016microwave-ablation}.
The target therapeutic temperature range ranges from $39-45^\circ$C for a more gentle effect on tumors and surrounding tissues, which is limited by the risk of vascular damage \cite{Paulides-etal-2020-review-hyperthermia}.
 
The advantages and disadvantages of microwave thermometry for medical diagnostic tasks are highlighted below.

\underline{Advantages} \cite{Stauffer-etal-2014brain-temperature, Osmonov-etal-2021coronavirus, Levshinskii-etal-2022venous, Goryanin-etal-2020microwave-radiometry-biomedical-studies, Groumpas-etal-2020brain-monitoring, mmwr-site}:
\begin{itemize}
\item [$\nearrow$] non-invasive method;
\item [$\nearrow$] very fast temperature measurement;
\item [$\nearrow$] inexpensive method;
\item [$\nearrow$] no contraindications;
\item [$\nearrow$] no restrictions on the procedure frequency;
\item [$\nearrow$] it is possible to measure both the thermodynamic temperature $T$ and local changes in the electromagnetic characteristics of the biological tissue (primarily the electrical conductivity), since MWR measures the brightness temperature $T_B$ by the electric field;
\item [$\nearrow$] the device for measuring brightness temperature is portable system.
\end{itemize}

\underline{Disadvantages} \cite{mmwr-site, Sedankin-Chupina-Vesnin-2018}:
\begin{itemize}
\item [$\searrow$] low accuracy of building temperature fields compared to the resolution of structures when using ultrasound, tomography, mammography, magnetic resonance elastography;
\item [$\searrow$] poor spatial error in measuring the brightness temperature in the plane and along the depth of the tissue;
\item [$\searrow$] 
The MWR method determines only the brightness temperature $T_B$, which requires additional data processing to relate to the real thermodynamic temperature $T$ and is model-dependent;
\item [$\searrow$] 
restrictions on the air temperature in the room where measurements are taken.
\end{itemize}

The main goal of the paper is to develop a method for diagnosing breast cancer based on machine learning algorithms using thermometric data, both real medical measurements and results of simulation modeling of MWR examinations. The main highlight of our approach is the integration of these two data sources into one combined dataset.
We combine spatial distributions of surface (IR thermography) and deep (MWR) temperatures for $M^{(real)}$ real patients (Subsection \ref{subsec:SUBD}) and $M^{(sim)}$ model patients, for which similar temperature distributions are calculated based on models of the dynamics of thermal and radiation fields inside a biological tissue with different internal characteristics.
3D breast models differ both in their internal geometric structure (Subsection \ref{subsec:3D-reconstruction}) and in the sets of physical parameters that determine the thermal and electromagnetic properties of various biological components (Subsection \ref{subsec:Electrical-thermal-characteristics}).
Such a combined sample is larger, most-robust and contains fewer errors and artifacts than only the results of real medical measurements.
Thus, Section \ref{sec:Materials-Methods} is devoted to the description of computer models needed to study the dynamics of thermal and electromagnetic fields inside a biological tissue.
We also describe the results of thermometric data processing in Section \ref{sec:result}, based on machine learning algorithms and neural networks.

Anticipating the presentation of our research results in Sections \ref{sec:Materials-Methods} and
\ref{sec:result}, we give a brief overview  of the achievements in the use of microwave radiation, both in medicine in general, and in solving the problem of improving the efficiency of breast oncology diagnostics (Subsections \ref{subsec:1-common} -- \ref{subsec:1-breast}).

\subsection{Application Fields of Microwave Radiation in Medicine}\label{subsec:1-common}

\begin{figure}[h!]
\begin{center}
  \includegraphics[width=0.6\hsize]{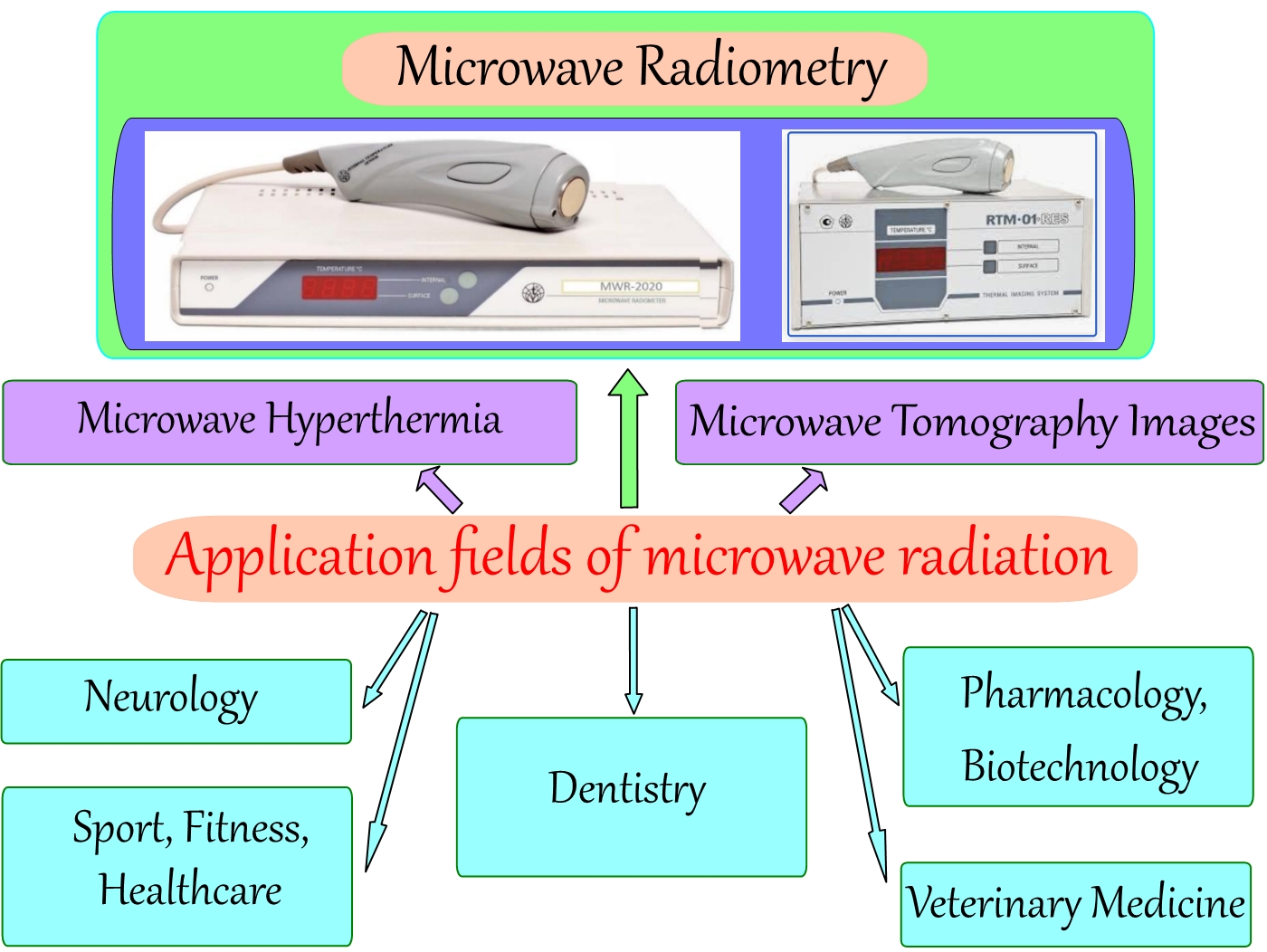}
\end{center}
\caption{
Some applications of microwave in medicine. 
 \label{fig:SchemeUseRTM}}
\end{figure}

The use of microwave radiation for various applications in medicine has a history of more than 80 years (See \cite{Pennes-1948thermal-model} and references there-in).
Figure~\ref{fig:SchemeUseRTM} demonstrates the areas of this kind of application of this electromagnetic range. 
 
In addition to diagnosing various diseases, we will point out applications in the field of sports, fitness and general healthcare monitoring. The development of microwave ablation methods as a clinical tool for the oncology treatment has great prospects \cite{Paulides-etal-2020-review-hyperthermia, Witte-Tamimi-2022thermo-therapy, Moon-Brace-2016microwave-ablation}.
A related use is associated with radiofrequency vein ablation for surgical purposes.
A high therapeutic effect is given by physiotherapeutic practice based on microwave to create anti-inflammatory, anti-allergic, immunostimulating and antispasmodic effects.
MW allows you to eliminate cosmetic skin defects, such as keloid scars.
Efficacy of microwave sterilization in dentistry are confirmed \cite{Sowmya-etal-2022Microwave-Sterilisation}.
The properties of passive microwave radiation are determined by complex biochemical and biophysical processes inside proteins, cells and organs as a whole with the release of energy, which makes MWR one of the tools in pharmacology at the stages of both preclinical and clinical studies \cite{Goryanin-etal-2020microwave-radiometry-biomedical-studies}.

The microwave radiometry has certain advantages for rapid mass diagnostics in comparison with magnetic resonance imaging (MRI) \cite{Liang-etal-2022magnetic-resonance-imaging}, mammographic screening \cite{Gastounioti-etal-2022mammographic-review}, ultrasound \cite{Akatsuka-etal-2022ultrasound},  thermoacoustic tomography \cite{Chi-etal-2022thermoacoustic-tomography}.
Magnetic resonance elastography (MRE) of the breast provides new opportunities as further development of ultrasound diagnostics to determine the biomechanical properties of tissues \cite{Bohte-etal-2018MRE}. Machine learning algorithms can improve both the accuracy and reliability of MRE for cancer diagnosis (See review \cite{Mao-etal-2022Elastography} and references there-in).
A separate approach to cancer diagnosis is tumor markers (biomarkers), which, as active substances, indicate the cancer presence  \cite{Brown-etal-2022statist, Hong-etal-2022Tumor-Markers}. The use of biosensors greatly facilitates the detection of relevant biomarkers compared to more traditional methods of searching for such breast tumor markers based on immunoassay, proteomics and other methods of molecular biology \cite{Hong-etal-2022Tumor-Markers}.
Above we indicated a number of advantages of MWR for diagnostics (non-invasiveness, low cost and short duration of the procedure, the absence of contraindications and restrictions on the measurements frequency, etc.).
Additionally, it should be noted that MWR gives sufficiently high sensitivity and specificity at the same time.
Therefore, MWR can be a tool for initial diagnosis of breast cancer, for which the earliest possible detection is a key factor in successful treatment.

\begin{figure}[h!]
\begin{center}
    \includegraphics[width=0.85\hsize]{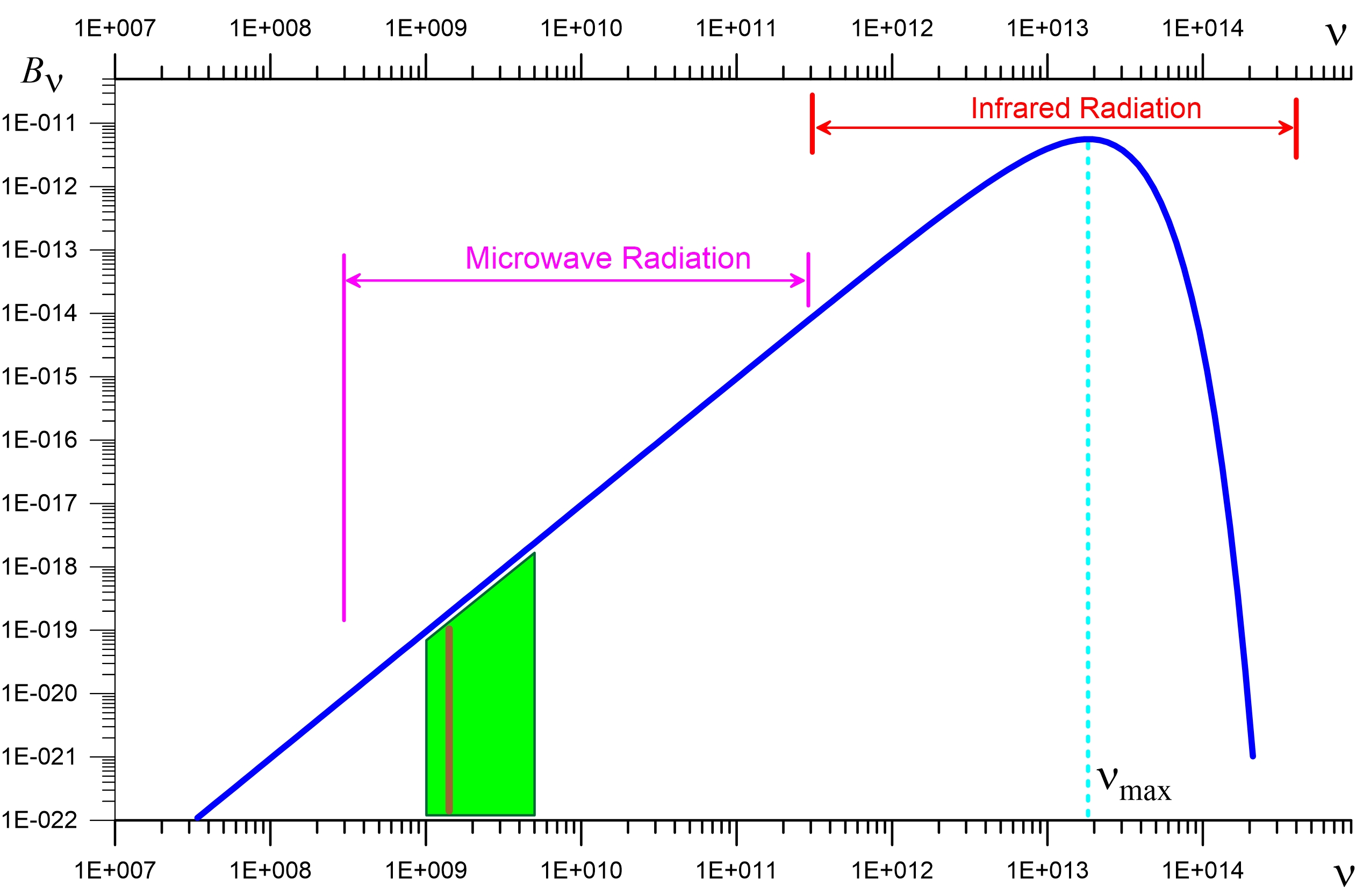}
\end{center}
\caption{
The spectral density of electromagnetic radiation $B_\nu$ at temperature $T=37^{\circ}$C (blue line): the left axis is the dependence $B_\nu(\nu; T)$, $\nu$ is the radiation frequency ($[\nu] = $Hz). Magenta line shows the range for microwave radiation. Other colored lines are explained in the text. 
 \label{fig:Spectr}}
\end{figure}

\subsection{Diagnostics Based on Microwave Radiometry}

Figure \ref{fig:Spectr} shows the spectral energy density of passive radiation ($B_\nu$) at body temperature, which is typical of the human core within $T_{core}=(37-38)^{\circ}$C at rest.
The maximum value of $B_\nu$ is reached at the frequency $\nu_{\max} \simeq  1.82\cdot 10^{13}$~Hz, which corresponds to the wavelength $\lambda_{\max} =1.64\cdot 10^{-5}$~m.
Instruments use different MW bands, which are marked with vertical lines in this figure. For example, the broadband radiothermometer RTM --- 01-RES can operate in the range of $1 - 5$\,GHz depending on the type of radio sensor (the green bar in Figure \ref{fig:Spectr}), the radiometer's small antenna (approximately 2~{cm}) provides reliable measurements in the range of $1.4 - 1.427$\,GHz from depths up to 2.3~{cm} (See the brown line) \cite{Yazdandoost-2021non-invasive-temperature}.
Antenna-applicator and sensor measure the internal (microwave) and skin (infrared) temperature at one point simultaneously in about 3 seconds (including the processing time).

The intensity of microwave radiation in the operating range is $10^{6}-10^{8}$~times less than in the IR range (See Figure \ref{fig:Spectr}), which gives us skin temperature based on IR thermography. Therefore, there is a direct dependence of the radiation intensity on temperature in the microwave range in accordance with the Rayleigh-Jeans formula 
\begin{equation}\label{eq:Bnu}
B_\nu= \frac{k_B \nu^2 }{ c^2}\cdot T \,,    
\end{equation}
where $k_B$ is the Boltzmann constant, $c$ is the speed of light. The linear dependence of $B_\nu$ on $T$ underlies the temperature measurement due to the proportionality between $T_B$ and $B_\nu$ under a narrow operating frequency range.

\begin{figure}[h!]
\begin{center}
    \includegraphics[width=0.9\hsize]{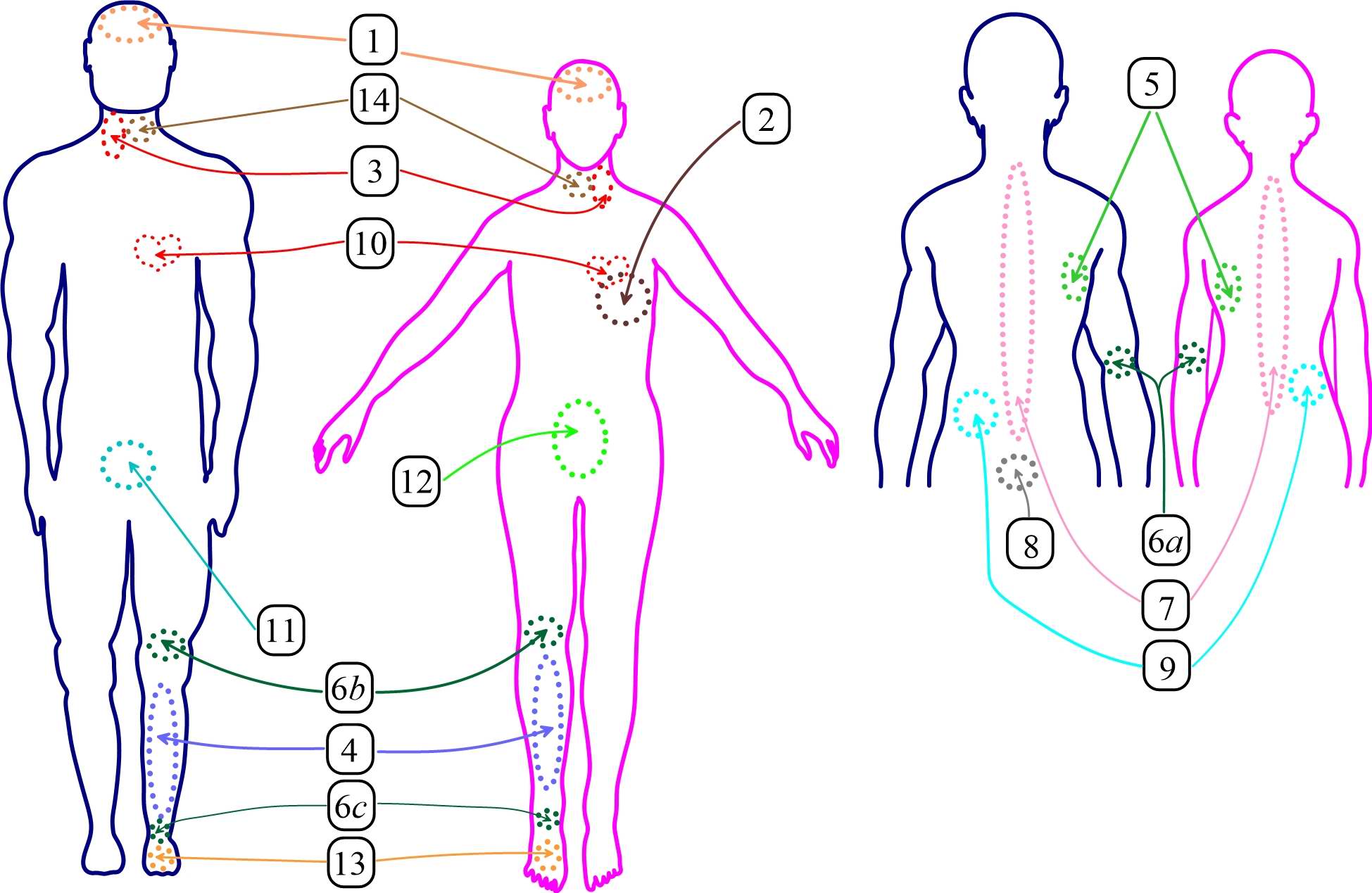}
\end{center}
\caption{\label{fig:MedicalUsesRTM}
The MWR method has been successfully applied to the following organs, tissues and diseases:
\textit{1} --- \mbox{brain \cite{Robert-etal-1979Millimeter-wave-thermography, Scapaticci-etal-2012microwave-imaging-brain, Shevelev-etal-2022brain, Leushin-etal-2022radiothermograph-Vesnin};} \textit{2} --- breast \cite{Barrett-etal-1977breast-cancer-microwave, Polyakov-etal-2021ourVest, Germashev-2022Cyber-Physical-System, Losev-Svetlov-2022}; \textit{3} --- carotid arteries \cite{Athanasiadi-etal-2022carotid-artery};  \textit{4} --- venous diseases of the lower extremities \cite{Levshinskii-etal-2022venous}; $\textit{5}$ --- coronavirus infection (Covid-19) \cite{Osmonov-etal-2021coronavirus};   \textit{6} --- arthritis of the large joints (elbow ($6a$), knee ($6b$), ankle ($6c$) \cite{Laskari-etal-2020-arthritis, Zampeli-etal-2013rheumatoid arthritis-knee}, rheumatoid arthritic knee joints \cite{Edrich-Smyth-1978Arthritis-thermography}, sacroiliac joints \cite{Laskari-etal-2018}; $\textit{7}$ --- spine arthritis \cite{Tarakanov-etal-2021spine}; $\textit{8}$ --- acute inflammatory diseases of the prostate
 \cite{Kaprin-etal-2019Urological-Diseases}; $\textit{9}$ ---  kidney disease \cite{Kaprin-etal-2019Urological-Diseases, Arunachalam-etal-2010bladder}; 10 --- cardiovascular diseases \cite{Benetos-etal-2021Carotid-artery}; 11 --- bladder \cite{Arunachalam-etal-2010bladder}; 12 --- pathology of the female reproductive system \cite{Chupina-Sedankin-Vesnin-2019development-medical-equipment}; 13 ---  diabetic foot ulceration pathology \cite{Spiliopoulos-etal-2017arterial-disease-diabetic}; 14 --- thyroid \cite{Vetshev-etal-2006thyroid-diseases}.
}\end{figure}

The diagnostic capabilities of microwave radiometry extend to a significant number of diseases of various organs/tissues (Figure \ref{fig:MedicalUsesRTM}).
The original studies were on the brain \cite{Robert-etal-1979Millimeter-wave-thermography, Scapaticci-etal-2012microwave-imaging-brain}, arthritis of various joints \cite{Edrich-Smyth-1978Arthritis-thermography}, breast \cite{Barrett-etal-1977breast-cancer-microwave}.
Since many diseases are associated with inflammatory processes, a local increase in internal temperature (at a depth of several centimeters) and the corresponding specific gradients of temperature fields can be detected by MWR measurements.
It is noteworthy that this method allows to reduce the number of errors and increase the informative features of X-ray and ultrasound diagnostic methods in the early stages of acute inflammatory diseases of the kidneys and prostate \cite{Kaprin-etal-2019Urological-Diseases} (See Figure~\ref{fig:MedicalUsesRTM}, pointers 8 and 9).

Brain activation due to local changes in blood flow during the metabolic activity of neurons is accompanied by changes in temperature and electrical conductivity, which can be tracked by real-time MWR \mbox{measurements \cite{Groumpas-etal-2020brain-monitoring}.} The microwave radiometry can provide additional clinical monitoring of deep temperatures during surgical operations on the brain, since lowering the temperature of the operated organs significantly increases the risk of pathological consequences \cite{Stauffer-etal-2014brain-temperature}. Brain activation due to local changes in blood flow during the metabolic activity of neurons is accompanied by changes in temperature and electrical conductivity, which can be tracked by real-time MWR monitoring \cite{Groumpas-etal-2020brain-monitoring}. Traumatic brain injuries (TBI) alter the temperature distribution due to disruption of normal blood circulation, and special studies show the ability of MWR to fix even small TBI \cite{Shevelev-etal-2022brain}. Separately, we single out non-invasive radiometric sounding for thermal monitoring of deep brain temperature in infants \cite{Hand-etal-2001deep-brain-temperature}.

Some joint diseases are accompanied by inflammatory processes, which makes MWR an effective tool for the early diagnosis of arthritis in various joints \cite{Goryanin-etal-2020microwave-radiometry-biomedical-studies, Laskari-etal-2018}. Local temperature changes in the pathophysiology of arthritis and arthrosis reflect inflammation even in the absence of clinical signs and can be diagnosed by MWR before pain occurs \cite{Laskari-etal-2020-arthritis}. The review \cite{Schiavon-etal-2021Review-Joint-Diseases} shows the effectiveness of local temperature measurement methods for a wide range of rheumatic diseases.

Measurements by the method of multifrequency three-dimensional (3D) radiothermogra\-phy are promising and are aimed at determining the depth and temperature of a cancerous tumor~\cite{Sidorov-etal-2021-3D-Visualization-Heat-Field}. The microwave range used makes it possible to determine the temperature at a depth of $2-6$ cm, depending on the biological tissue properties and frequency.
New technical possibilities for multifrequency measuring the brightness temperature require appropriate mathematical models for a more accurate determination of both the localization and the thermodynamic temperature of the studied areas. Intracavity measurements of brightness temperature using new generations of antennas through natural cavities make it possible to determine the dynamics of three-dimensional $T_B$ distributions \cite{Sedankin-etal-2021Intracavity-Thermometry-Medicine}. We also point out the possibility of continuous monitoring of the internal temperature of the brain in newborn infants using a multi-frequency microwave radiometer during cooling of the brain after hypoxia-ischaemia \cite{Hand-etal-2001deep-brain-temperature}.

Measurements of temperature difference along the carotid artery underlie the method of presymptomatic diagnosis of myocardial dysfunction, coronary heart disease, peripheral arterial disease (See pointer 3 in Figure \ref{fig:MedicalUsesRTM}) \cite{Athanasiadi-etal-2022carotid-artery}, and also provide control of the treatment process \cite{Benetos-etal-2021Carotid-artery}. The development of a new instrument base significantly expands the areas of medical application of MWR, including the ability to monitor the pelvic organs \cite{Sedankin-Gudkov-2019Pelvic-Organs}. Intracavitary sensors for monitoring intravaginal temperature are complete tools for continuous monitoring of important physiological processes \cite{Law-etal-2018brown-adipose-tissue}. Such approaches make it possible to quickly detect objective changes in the physiological state, including pregnancy planning \cite{Zaretsky-etal-2018arterial-blood-temperature}. Microwave sensors are already capable of detecting inflammation of the ovaries and cervix \cite{Chupina-Sedankin-Vesnin-2019development-medical-equipment}.
Microwave thermometry is not limited to clinical practice and is used in pharmacology, physiological research to create thermal maps of muscles and internal organs in sports and health monitoring \cite{Raiko-etal-2022temperature-measurements}.

\subsection{MWR method for detecting breast cancer}\label{subsec:1-breast}

Mass-produced medical devices RTM\,--\,01--RES with modifications (manufactured by RES, Ltd., Russia, \mbox{Moscow \cite{Sedankin-etal-2021Intracavity-Thermometry-Medicine},} See images of device in Figure \ref{fig:SchemeUseRTM}) are basis of medical diagnostics of both diseases of breast \cite{Polyakov-etal-2021ourVest, Germashev-2022Cyber-Physical-System} and other \mbox{organs \cite{Osmonov-etal-2021coronavirus, Levshinskii-etal-2022venous, Leushin-etal-2022radiothermograph-Vesnin}.} Our measurement data for temperature distributions for breast were obtained with this device, that provides an accuracy of about $\pm 0.2^\circ$C \cite{Sedankin-Chupina-Vesnin-2018}.

\begin{figure}[h!]
\begin{center}
    \includegraphics[width=0.7\hsize]{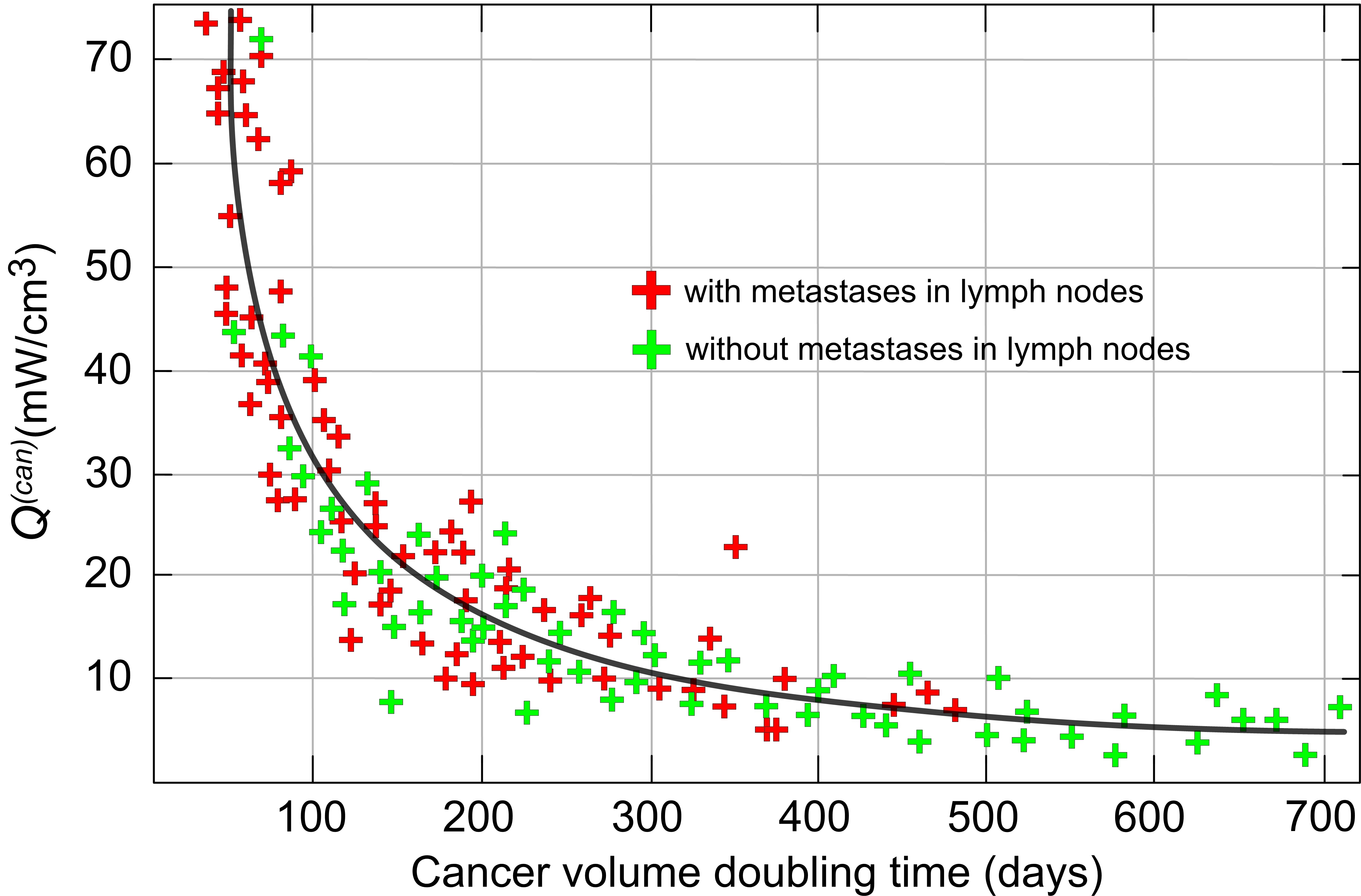}
\end{center}
\caption{\label{fig:cancer-energy}
Dependence of tumor specific heat release $Q^{(can)}$ on doubling time for 128 breasts, where tumor diameter is 0.4 cm\,$\le D \le $\,4 cm \cite{Gautherie-1982}.
}\end{figure}

The application of MWR is always based on combined use of IR thermography, which gives surface skin temperature distribution. Analysis of these two spatial temperature distributions underlies breast cancer diagnosis. Cancer formations are a local source heat additional inside tissue (Figure~\ref{fig:cancer-energy}) and lead to characteristic changes in temperature distribution $T_{IR}$ and $T_B$.

The traditionally used diagnostic methods (ultrasound, computer tomography, mam\-mography)
do not allow to effectively detect a tumor at an early stage, when its size is small and heat generated is low. One of the most promising methods for improving efficiency of mammographic screening and early differential diagnosis is MWR method. 

A separate area of breast cancer diagnostics based on MWR data is analysis of feature spaces \cite{Levshinskii-etal-2019politeh, Levshinskii-etal-2020-application-data-mining, LosevLevshinskii-2017-Math-Phys-Comput-Simul}.
These are built in form of various combinations of temperature differences measured at different points in breast.
Algorithms such as Random Forest, XGBoost, K-Nearest Neighbors, Support Vector Machine, cascade-correlation neural network, deep neural network and convolutional neural network, decision trees and naive Bayesian classifier are used to classify thermometric data \cite{Losev-etal-2022Machine-Learning, Levshinskii-etal-2020-application-data-mining, Zuluaga-Gomez-etal-2019thermography-breast-cancer}.
This causes an increase in values sensitivity and specificity for cancer detection and a decrease in errors in diagnoses in early stages of tumor growth \cite{Losev-Svetlov-2022, Levshinskii-etal-2020-application-data-mining, Vesnin-etal-2017-Modern-microwave-thermometry}.
In \cite{GalazisVesninGoryanin-2019-application-artificial-intelligence}, deep neural network showed the best result on test set with a significant improvement over other machine learning algorithms.
However, disadvantage of neural networks is ``black box'' type structure, which does not allow interpretation of results \cite{Schonberger-2019-artificial-intelligence}.
Analysis of thermometric data should be part of expert diagnostic systems.
Results of classification should be presented in terms that allow justification of the diagnosis, which is extremely important for medical practice.

Medical examinations and analysis of MWR data revealed important features of breast cancer \cite{Levshinskii-etal-2020-application-data-mining, LosevLevshinskii-2017-Math-Phys-Comput-Simul, Kobrinskii-2008-intelligent-medical-systems}.
These include the following: increased value of thermal asymmetry between corresponding points of breast;  high temperature difference between separate points of breast with a tumor; high difference between nipple temperatures; increased nipple temperature compared to average temperature of breast, etc.

Analysis of feature spaces required development of special descriptive mathematical models of the patient's diagnostic state, on basis of which it is possible to effective classification models and justification the diagnostic result \cite{Levshinskii-2021-Chelyab}.
Classification algorithms are based on feature spaces such as temperature, thermometric features, 2nd, 3rd and 4th degree polynomial features \cite{Levshinskii-2021-Chelyab}. 
The use of some artificial intelligence algorithms allows increasing sensitivity value up to 0.892 and specificity value up to 0.813.

Let us point out some problems of MWR diagnostics and their solution in this article.
Medical measurement datasets do not contain quantitative characteristics of tumors.
There is only information about the diagnosis of ``cancer''.
Therefore, training of neural networks based on medical examination data takes into account only the fact of presence/absence of a disease (binary feature).

We propose to use numerical simulations results of thermal processes in breast models in addition to real medical measurements.
This allows you to use multidimensional quantitative characteristics of tumor instead of a binary feature (sick/healthy) in algorithm input layer.
The diagnosis of ``cancer'' based on medical measurements is made only on basis of validated additional studies, and datasets do not indicate presence of mild tumors, even if some patients have them.
Extending dataset with numerical simulation complements training samples with weak tumors as well.

\section{Materials and Methods}\label{sec:Materials-Methods}

Simulation of physical processes inside a biological tissue requires the use of realistic models that take into account their complex structure. The breast consists of a large number of biologically distinct components (skin, adipose tissue, lobe and lobules, areola, nipple, lactiferous sinus, lactiferous duct, muscle, ductule, subcutaneous fat pad, suspensory ligaments, lymph nodes, rectoral fat pad).
The procedure for constructing geometric models of the internal structure of the breasts is described in Subsection \ref{subsec:3D-reconstruction}.
The realization of one such structure is determined by the vector $\vec{\cal G}$.
Each biological component is characterized by its own set of physical characteristics (electrical conductivity, dielectric constant, thermal conductivity, specific heat release, heat capacity, moisture content, and others), which have some variation in values in different model patients.
The vector $\vec{\cal F}$ defines a set of physical parameters for one virtual patient (Subsection \ref{subsec:Electrical-thermal-characteristics}).

\subsection{Method for 3D Reconstruction of Multicomponent Tissue} \label{subsec:3D-reconstruction}

\begin{figure}[h!]
    \centering
    \includegraphics[width=0.65\hsize]{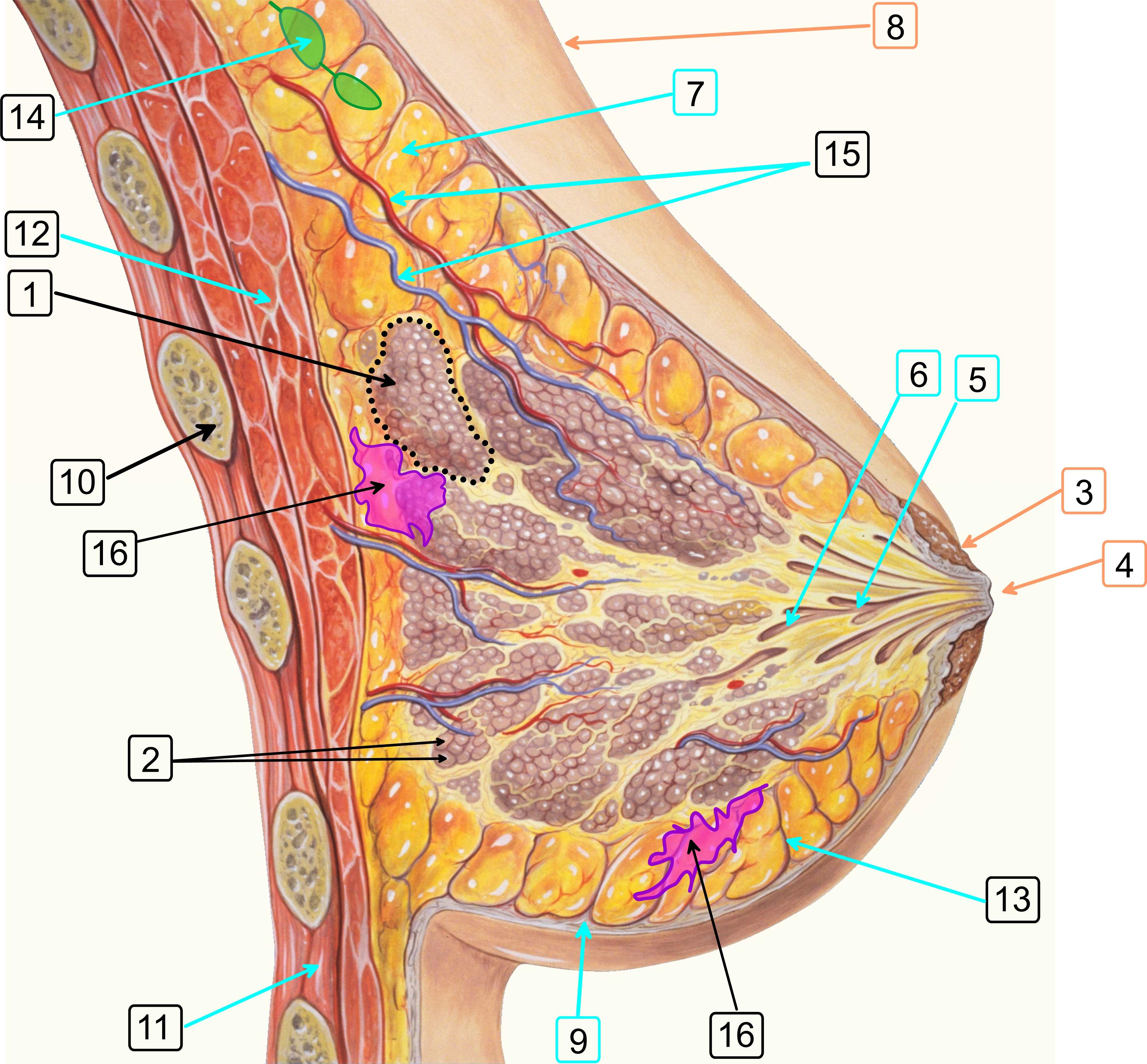}
    \caption{ 
The structure of the breast: 
1 --- the breast lobe; 
2 --- the lobules; 
3 --- the areola; 
4 --- the nipple; 
5 --- the lactiferous sinus; 
6 --- the lactiferous duct;  
7 --- the adipose tissue; 
8 --- the skin; 
9 --- the subcutaneous fat pad; 
10 --- the rib cage; 
11 --- the intercostal muscle; 
12 --- the pectoralis major muscle; 
13 --- the suspensory ligaments (of Cooper); 
14 --- the lymph node; 
15 --- the circulatory system (arterial and venous subsystems);  
16 --- the tumor.
    (The basis of medical illustration: Patrick J. Lynch, medical illustrator; C. Carl Jaffe).
    } \label{fig:3D-breast}
\end{figure}

The shape and size of the breasts vary in a fairly wide range, which affects the spatial distribution of temperature.
An equally important factor is the internal structure, which is determined by the location of various components of connective, muscle, epithelial tissues, as well as their size and geometric shape (Figure \ref{fig:3D-breast}).

The construction of geometric model of the internal structure of biological tissues can be based on 3D reconstruction using data from a repository of magnetic resonance imaging images, or high-precision layer-by-layer grinding of frozen biological tissues \cite{Khoperskov-etal-2017-3D-reconstruction}, which is full-color and provides resolution to 5 $\mu$m\, in contrast to MRI data with a resolution of approximately 0.5~{cm}.
An important disadvantage of these approaches is the small number of distinct instances with different internal structures. 
Therefore, these methods can only be of an auxiliary nature.

\begin{figure}[h!]
\begin{center}
    \includegraphics[width=0.75\hsize]{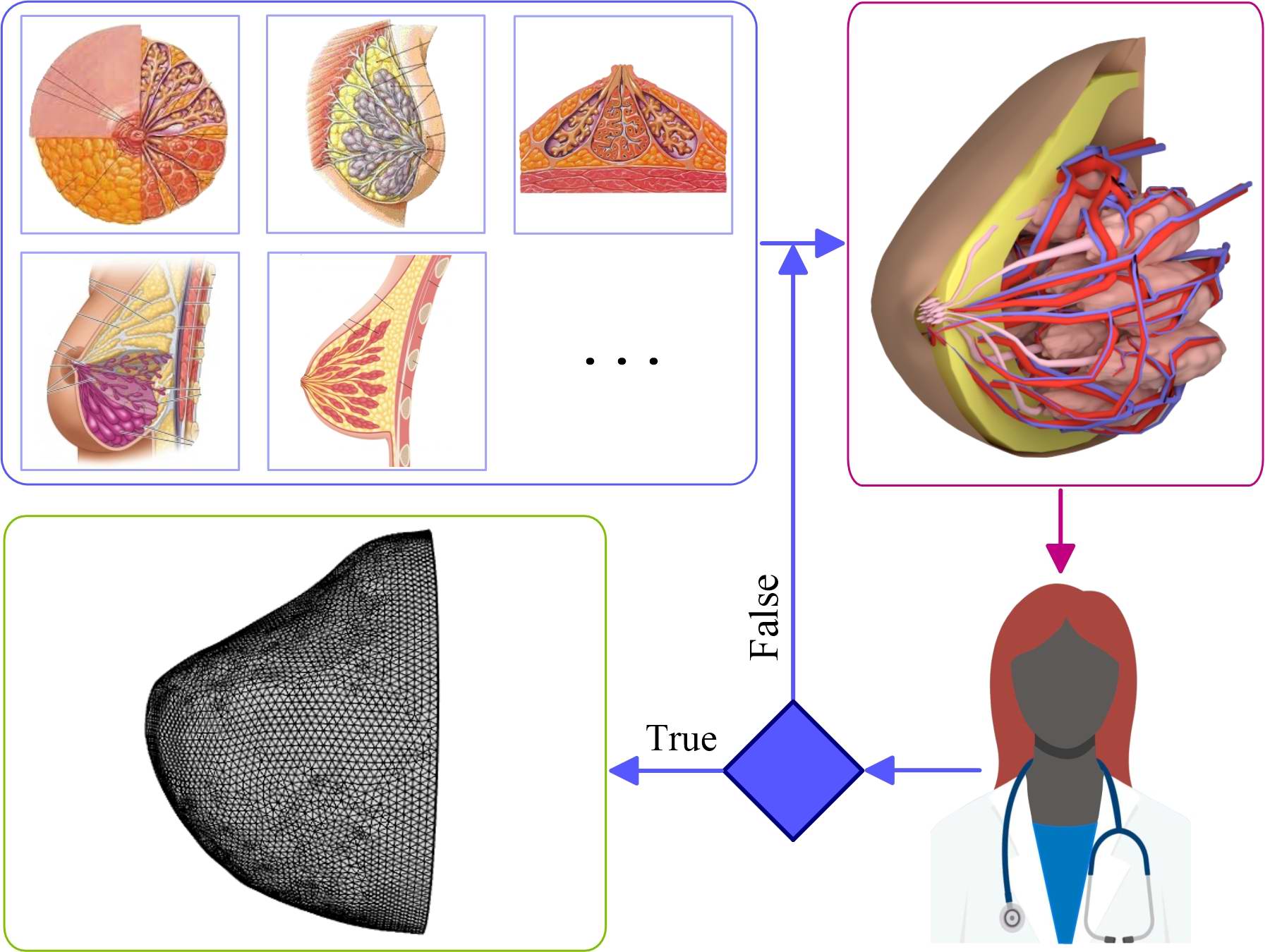}
\end{center}
\caption{\label{fig:Scheme2} General procedure for constructing 3D model of biological tissue.}
\end{figure}

We use an iterative method based on the complex application of data from medical atlases and expert recommendations from physicians, specialists in the field of breast anatomy. The initial data are anatomical images of the breast in various projections (Figure~\ref{fig:Scheme2}). Further, special software Blender allows you to build separate elements of a vector 3D model that define one or another component. Then the elements are combined into a single geometric model. The experts verify the model, pointing out the necessary adjustments until the model reaches a high quality, conveying all the key anatomical features of the organ under study. Three-dimensional computational grid is built at the final stage. Figure~\ref{fig:Scheme2} shows the scheme of the described algorithm. 
 We use the free and open-source software Blender v.2.82a (GNU GPL 3 license) to build 3D models of the internal structure of the breast

\begin{figure}[h!]
    \centering
    \includegraphics[width=0.99\hsize]{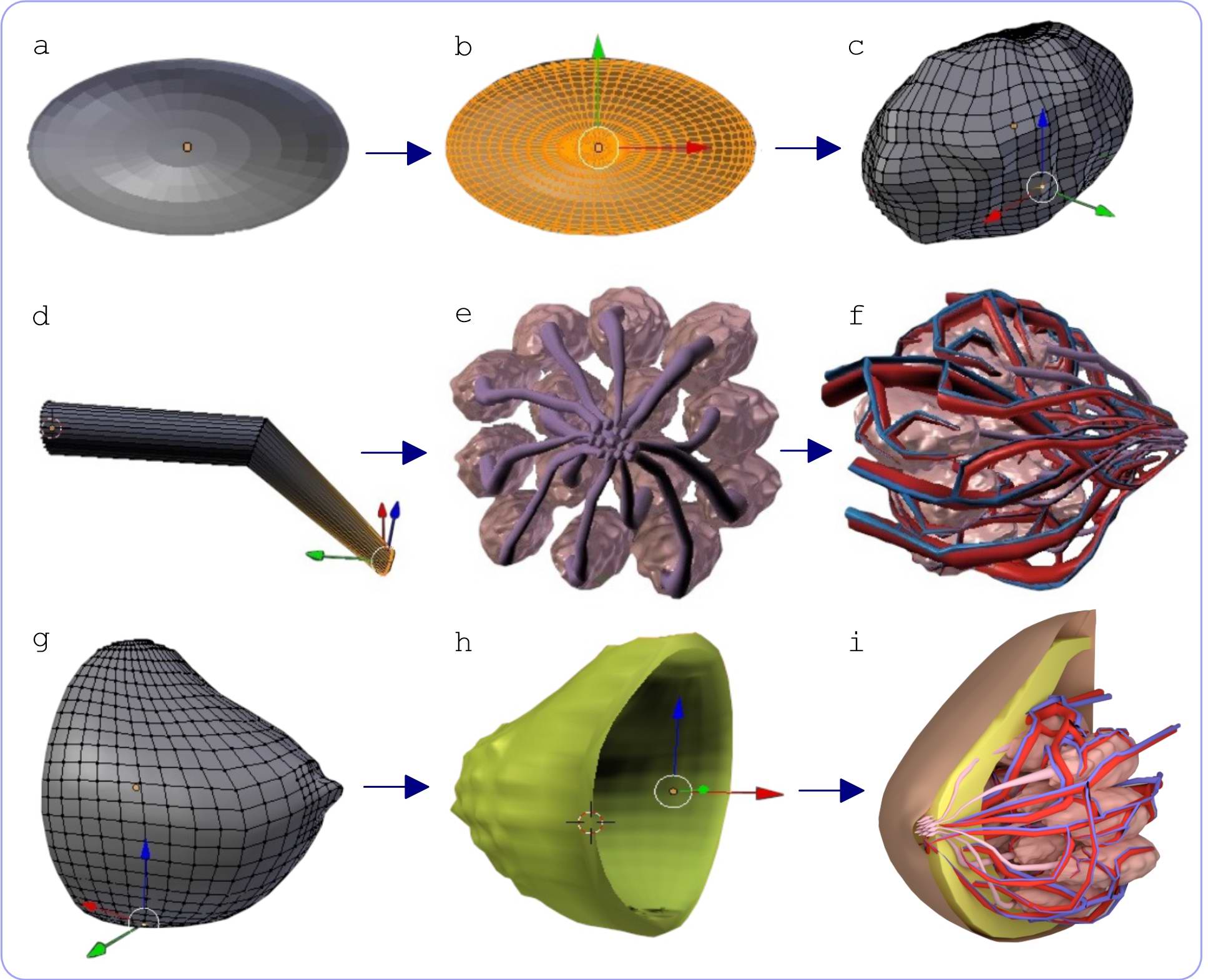}
    \caption{ Procedure to construct a 3D breast model of different components in Blender: steps for constructing a single breast lobe model (a -- c), system of the lactiferous sinus and breast lobes (d -- e), the blood flow includes arteries (red) and veins (blue)  (f), the skin (g), the subcutaneous fat pad (h), the section of the final 3D model of the breast with textures (i).
    } \label{fig:3D-modeling-algorithm}
\end{figure}

Models of individual components of the breast were created in the Blender software package (Figure \ref{fig:3D-modeling-algorithm}).
The figure shows examples of building 3D models of some components of the breast. 
Note that blood flows form the hierarchical circulatory system, including arteries, veins, arterioles, capillaries, venules, which have different diameters. 
We texture each 3D object just for visual representation (See Figure \ref{fig:3D-modeling-algorithm}\,i) and the texture is not taken into account further in the computational model. 
The script in Blender allows you to build the biocomponents of the model with a resolution of up to 0.01 cm. Given the individual variations of the internal structure of the models and the fact that the characteristic electromagnetic wavelength for the MWR method is about $\geq$ 2 cm, this resolution does not distort the final result of statistical processing.
Thus, we have the basic geometric model of the breast with a complex and realistic internal structure, which is determined by the following set of parameters
\begin{equation}\label{eq:def-G}
\vec{\cal G} = \left\{ g^{(1)}_1, g^{(1)}_2, \ldots, g^{(1)}_{k_1}, g^{(2)}_1, g^{(2)}_2, \ldots, g^{(2)}_{k_2}, \ldots ,  g^{(K)}_{1}, g^{(K)}_2, \ldots, g^{(K)}_{k_K} \right\} \,,     
\end{equation}
where $g^{(i)}_{j}$ is the $j$-th geometric characteristic (sizes, coordinates, orientation angles) for $i$-th object, the number of parameters $k_i$ differs for different biological components, obviously, $K$ is the total number of objects that form the breast model.
Some basic model is described by the vector $\vec{\cal G}_0=\big\{ \big(g^{(i)}_{j}\big)_0 \big\}$, which determines the average geometric characteristics.
Each geometric parameter has natural variation within $\pm \delta{g}^{(i)}_{j\,\max}$:
\begin{equation}\label{eq:limit-parameters}
   \big({g}^{(i)}_{j}\big)_0 - \delta{g}^{(i)}_{j\,\max}  \le {g}^{(i)}_{j} \le \big({g}^{(i)}_{j}\big)_0 + \delta{g}^{(i)}_{j\, \max} \,, 
\end{equation}
where $\delta{g}^{(i)}_{j\,\max}$ is agreed with the medical expert group. 
We choose determined parameter values
\begin{equation}\label{eq:gij-parameters}
{g}^{(i)}_{j} =  \big({g}^{(i)}_{j}\big)_0 + \delta{g}^{(i)}_{j\, \max} \cdot \xi \,,
\end{equation}
where $\xi \in [-1;1]$ is the random normal number. 

The result of such a procedure is sets of geometric models with different internal structures $\vec{\cal G}_m$ ($m=1,\ldots, M$), the number of which is $M = 600$.
Figure \ref{fig:3D-modeling-algorithm}\,i demonstrates the model's internal structure with the basic set of geometric parameters $\vec{\cal G}_0$. 

We are trying to construct a realistic internal structure of the breast in numerical models of the dynamics of thermal and radiation fields (See Subsections 2.2 and 2.3 below). This distinguishes our approach from traditional multilayer biotissue models \cite{Sedankin-etal-2018Mathematical, Zuluaga-Gomez-etal-2019thermography-breast-cancer, Ferreira-Numerical-analysis-2019, Mukhmetov-et-al-2021}, which use a sequential set of homogeneous layers.

\subsection{Electrical and Thermal Characteristics of Biological Tissues}\label{subsec:Electrical-thermal-characteristics}
Each biological component of the breast (See Figure \ref{fig:3D-breast}) has its own values of physical parameters.
The main characteristics include the thermal conductivity $\lambda$, the heat capacity $C$, the mass density of matter $\rho$, the heat source power density due to metabolic processes $Q^{(met)}$, the electrical conductivity $\sigma$, the permittivity $\varepsilon$.
Our analysis of these characteristics from the scientific literature is shown in Table \ref{tab:diapason_thermophysical_properties}), which allowed us to determine the ranges in which the values of such parameters vary. Moreover, the value of $Q^{(can)}$ for tumors varies widely at different stages of disease development (See Figure \ref{fig:cancer-energy}).
It should be emphasized that the electromagnetic parameters of various components may depend on the radiation frequency $\nu$ (the data in Table \ref{tab:diapason_thermophysical_properties} are given for the frequency $\nu = 1.5$ GHz).

\begin{table}[h!]
	\caption{The range of natural variability in the physical properties of the main components of the breast \cite{Gautherie-1982, Chen-2021, Bhowmik-2014, Stafford-1951, Sudharsan-2001, Bowman-1975, Iljaz-2019, Gordon-1976, Gomboc-2021, Hensel-1952, Lipkin-1954, Hossain-2016, Hatwar-2017, Bezerra-2013}. 
 }
	\begin{center}
		\footnotesize
		\begin{tabular}{lcccccc}
			\hline
			& $\lambda$, W/(m$\cdot$K)\ & $C$, J/(kg$\cdot$K) & $\rho$, kg/m$^3$ & $Q^{(met)}$, W/m$^3$ & $\sigma$, S/m & $\varepsilon$ \\ \hline
			Skin           & 0.21--0.54                     & 3391--3690              & 1180--1215        & 380--410             & 1.1--2.4        & 40--50         \\
			Muscles          & 0.4--0.56                      & 3421--3790              & 1070--1100        & 675--690             & 0.44--0.7       & 54--56         \\
			Fat            & 0.18--0.34                     & 2348--2690              & 900--915          & 356--370             & 0.03           & 4.4--6         \\
			Blood      & 0.45--0.6                      & 3800--4200              & 1046--1058        & 0                   & 0.9--1.2        & 64--85         \\
			Glands         & 0.4--0.5                       & 3700--3790              & 1035--1041        & 450--610             & 0.56--0.61      & 10.6--12       \\
			Connective tissue & 0.44--0.5                      & 3340--3400              & 1006--1020        & 604--620             & 0.3--0.36       & 38--40         \\
						Cancers        & 0.45--0.58                     & 3710--3800              & 1045--1054        & 3000--71000          & 0.79--1.5       & 42--50         \\ \hline
		\end{tabular}
	\end{center}
	\label{tab:diapason_thermophysical_properties}
\end{table}

The set of physical parameters of the breast model is determined by the vector $\vec{\cal F}$, the meaning of which is similar to the vector $\vec{\cal G}$ (See (\ref{eq:def-G})):
\begin{equation}\label{eq:def-F}
\vec{\cal F} = (
\vec{F}_\lambda ,
\vec{F}_C ,
\vec{F}_\rho ,
\vec{F}_Q ,
\vec{F}_\sigma ,
\vec{F}_\varepsilon,
T_{air},
T_{core}
) \,,
\end{equation}
where the dimension of each of the $\vec{F}_{\lambda,\,\ldots,\,\varepsilon}$ vectors is equal to the total number of objects ($K$) that make up the entire model, $T_{air}$ is the air temperature in the room during measurements, $T_{core}$ is the human core temperature (See subsection 2.3 for detailed discussions).  
Each physical quantity is characterized by a corresponding possible deviation (See Table \ref{tab:diapason_thermophysical_properties}), for example, the heat capacity of fat varies within $\delta{C} = \pm 171$\,\,J/(kg$\cdot$K).
We distribute the values of the components of the vector $\vec{\cal F}$ according to the normal law similar to the geometric characteristics (See Subsection \ref{subsec:3D-reconstruction}, formulas (\ref{eq:limit-parameters}, \ref{eq:gij-parameters})), which allows generating samples of models with random characteristics, each of which is determined by its own tuple $\left\langle \vec{\cal G}, \vec{\cal F} \right\rangle$.
Physical characteristics vector
\begin{equation}\label{eq:def-F}
\vec{\cal F} = \left\{ f^{(1)}_1, f^{(1)}_2, \ldots, f^{(1)}_{k_1}, f^{(2)}_1, f^{(2)}_2, \ldots, f^{(2)}_{k_2}, \ldots ,  f^{(K)}_{1}, f^{(K)}_2, \ldots, f^{(K)}_{k_K}, f^{(air)}, f^{(core)} \right\} \,,     
\end{equation}
is defined similarly to the vector $\vec{\cal G}$ (See (\ref{eq:def-G})), $f^{(i)}_{j}$ is the $j$-th physical characteristic (thermal conductivity, heat capacity, mass density, heat release rate, electrical conductivity, dielectric constant) for $i$-th object, $f^{(air)}=T_{air}$, $f^{(core)}=T_{core}$.
We write for physical parameters as in the formula (\ref{eq:gij-parameters}):
\begin{equation}\label{eq:fij-parameters}
{f}^{(i)}_{j} =  \big({f}^{(i)}_{j}\big)_0 + \delta{f}^{(i)}_{j\, \max} \cdot \xi \,,
\end{equation}
where $\big({f}^{(i)}_{j}\big)_0$ and $\delta{f}^{(i)}_{j\, \max}$ are determined by data from Table \ref{tab:diapason_thermophysical_properties}. 

The result of generating 3D models for subsequent numerical simulations is about 2000 models, which are mixed in various proportions with medical measurement data and for which the spatial distributions of thermodynamic ($T$) and brightness ($T_B$) temperatures are calculated.

\subsection{Models of the Dynamics of Thermal and Radiation Fields}

The computational model should reproduce the process of measuring the brightness temperature $T_B$ inside the tissue.
The value of $T_B$ is determined by the distributions of both the thermodynamic temperature and the electric field.
Therefore, both of these quantities must be calculated self-consistently inside a multicomponent biological tissue with  complex structure on very different scales from $10-20$ cm to approximately {0.01~cm}.

Heat dynamics is determined by the heat conduction equation with different sources \cite{Pennes-1948thermal-model, Polyakov-etal-2017zamech, Gonzalez-2007Thermal-simulation, Vesnin-Sedakin-2012antenna-applicator, Polyakov-2020, Zuluaga-Gomez-etal-2019thermography-breast-cancer} 
\begin{equation}\label{Main-model}
	\rho (\vec{r})\, C(\vec{r}) \frac{\partial  T(\vec{r},t)}{\partial t} = \nabla \left( \lambda(\vec{r}) \nabla T \right) +Q^{(met)}(\vec{r},t)+ Q^{(bl)}(\vec{r}) + Q^{(can)}(\vec{r}) \,, 
\end{equation}
where $\rho$ is the mass density, $C$ is the  heat capacity of tissue, $T$ is the thermodynamic temperature, $\lambda$ is the thermal conductivity coefficient of biological tissue, $\vec{r} = \{x, y, z\}$, $\nabla$ is the nabla operator. 
We distinguish the following sources of heat source power density due to metabolic processes, produced by the metabolic processes in tissues ($Q^{(met)}$), the blood flows ($Q^{(bl)}$), the cancerous tumors ($ Q^{(can)}$), $[Q] =$W$\cdot$m$^{-3}$ \cite{Levshinskii-etal-2019politeh, Polyakov-etal-2021ourVest, Wren-2004Bio-heat Equation, Avila-Castro-etal-2017thermographic-simulator}.

\begin{figure}[h!]
\begin{center}
    \includegraphics[width=0.55\hsize]{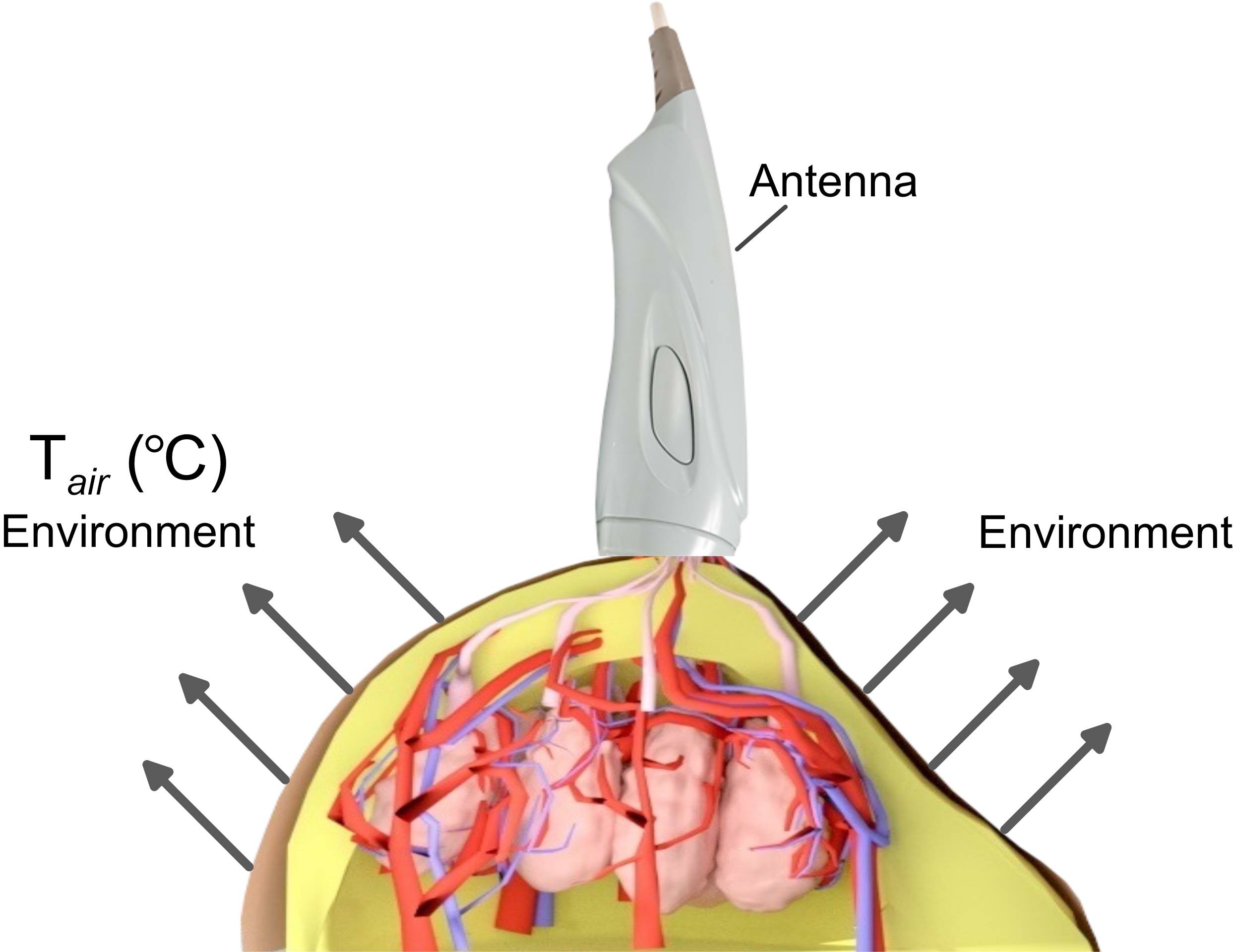}
\end{center}
\caption{\label{fig:task_geometry} Scheme for modeling measurements of the brightness temperature and the IR temperature of the skin.
}\end{figure}

The distinctive feature of our approach is taking into account the spatial heterogeneity of physical parameters at different scales in the realistic multicomponent biological tissue, including hierarchical circulatory system.
This can significantly change the distributions of brightness and IR temperatures in our model compared to traditionally used multilayer models \cite{Bardati-2008, Sedankin-etal-2018Mathematical}.
The boundary conditions between the biological tissue (skin) and the environment are based on the continuity of the energy flux (Figure~\ref{fig:task_geometry})
\begin{equation} \label{boundary}
	\lambda(\vec{r})\,\left(\vec{n}\cdot\nabla T(\vec{r},t)\right) = h\cdot (T_{air}-T(x, y, z, t)),
\end{equation}
where $\vec{n}$ is the normal vector to the boundary of the interface ``biological tissue~-- environment'', $h$ is the heat transfer coefficient (W/m$^2 \cdot K$), $T_{air}$ is the ambient temperature.
Measurements of internal and IR temperatures are carried out approximately 15~minutes after the start of the medical examination, when the patient gets used to the air temperature in the room~\cite{Polyakov-etal-2017zamech}.

The ambient temperature is a variable parameter when creating the dataset $\left\langle \vec{\cal G}, \vec{\cal F} \right\rangle$, so $T_{air}$ is set similarly to other physical characteristics of biological tissues with $T_{air\,0} = 23^{\circ}$C, $\delta T_{air\,\max} = 2^{\circ}$C in accordance with the actual conditions of medical measurements.
The second boundary condition is set for the pectoralis major muscle (See label 12 in Figure~\ref{fig:3D-breast}), for which the temperature $T_{core}$ is fixed.
The value of $T_{core}$ is also variable when creating the vector of physical characteristics ${\cal F}$, so that $T_{core\,0} = 37.5^{\circ}$C, $\delta T_{air \,\max} = 0.5^{\circ}$C (See (\ref{eq:fij-parameters})).
 
Our numerical integration of the non-stationary equation (\ref{Main-model}) together with (\ref{boundary}) makes it possible to construct a quasi-stationary solution in time interval of approximately 15 minutes, which is consistent with the medical examination technique.

Calculations of the electric field distributions in the breast are based on the numerical integration of Maxwell's equations:
\begin{equation}\label{eq-Makswell}
	\frac{\partial \vec{B}}{\partial t} + \nabla\times\vec{E} = 0 \,, \quad \frac{\partial \vec{D}}{\partial t} - \nabla\times\vec{H} = 0 \,,\quad \vec{B}=\mu\vec{H}\,,\quad \vec{D}=\varepsilon\vec{E} \,,
\end{equation}
where $\vec{E}$ is the electric field, $\vec{B}$ is the magnetic field, $\varepsilon$ is the permittivity, $\mu$ is the magnetic permeability.  

The result of the numerical integration of the equations (\ref{eq-Makswell}) is the stationary distribution of $\vec{E}(\vec{r})$ inside the biological tissue, which is necessary to calculate the brightness temperature $T_B$.

The inhomogeneity of the permittivity is a significant factor affecting the spatial distribution of the electric field, when $\varepsilon$ can change by an order of magnitude at the boundary of two components (for example, ``fat -- muscles''), which noticeably changes the vector $\vec{E}$.
The spatial distribution of the electric field in the monochromatic limit is described by the Helmholtz equation
\begin{eqnarray}\label{eq-Gelmgoltz}
	\Delta\, \vec{E}+\frac{\omega^2}{c^2}  \varepsilon \,\vec{E} = - \nabla \left( \vec{E}\cdot \vec{\nabla} (\ln\,\varepsilon) \right) \,,
\end{eqnarray}
where $\varepsilon(x,y,z;\nu)$ is the permittivity, $c$ is the speed of light in vacuum, $\omega=2\pi \nu$. 
The right side in the equation (\ref{eq-Gelmgoltz}) clearly shows the influence of the inhomogeneity of dielectric properties in a biological tissue.

\subsection{Calculation of Brightness Temperature in Biological Tissues}

The brightness temperature is determined by the volume integral \cite{Levshinskii-etal-2019politeh, Polyakov-etal-2021ourVest, Arunachalam-etal-2010bladder, Vesnin-Sedakin-2012antenna-applicator, Bardati-2008, Sedankin-etal-2018Mathematical}
\begin{equation}\label{eq-TBend}
	T_B =\int\limits_{V_b} \Omega(x,y,z;\nu)\, T(x,y,z)\,dV \,,
\end{equation}
where $V_b$ is the volume in which the microwave radiation is formed and then intercepted by the antenna (See Figure~\ref{fig:task_geometry}). 
The quantity
\begin{equation}\label{weight}
	\Omega=\frac{P_d(x,y,z;\nu)}{\int_{V_b} P_d dV}
\end{equation}
under the normalization condition 
$$
\int_{V_b} \Omega\,dV = 1 
$$
determines the weight function in terms of the electromagnetic energy density 
\begin{equation}\label{Pd}
	P_d = \frac{1}{2}\,\sigma(x,y,z;\nu)\cdot |\vec{E}(x,y,z;\nu)|^2,
\end{equation}
where $\sigma$ is the electrical conductivity.
It is important to emphasize that the electromagnetic characteristics of biological tissues depend on the frequency of microwave radiation.
{A detailed description of the numerical implementation of the mathematical model is discussed in the works \cite{Polyakov-etal-2021ourVest, Polyakov-Khoperskov-Svetlov-2017}.}

\begin{figure}[h!]
\begin{center}
  \includegraphics[width=0.95\hsize]{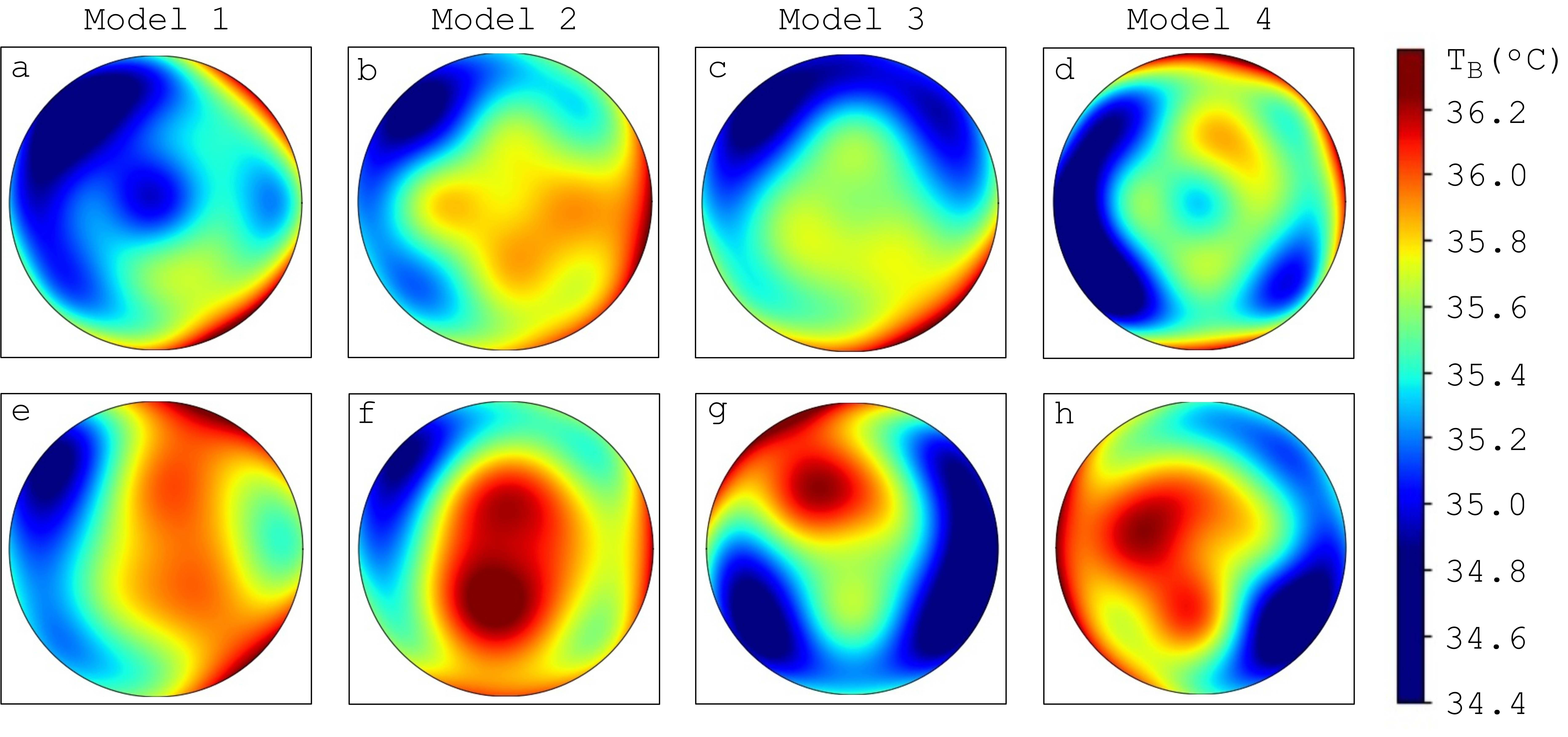}
\end{center}
\caption{
Brightness temperature distributions based on the results of simulations using a modified measurement scheme for four models (1 --- a, e; 2 --- b, f; 3 --- c, g; 4 --- d, h): top row --- without tumor, bottom row --- with tumor. Tumor parameters: $Q^{(can)} = 35 400$\,W$\cdot$m$^{-3}$, $R=1$\,cm.  
 \label{fig:bright-temperature-8fig}}
\end{figure}

Figure~\ref{fig:bright-temperature-8fig} internal temperature distributions for models with different sets of parameters $\left\langle \vec{\cal G}, \vec{\cal F} \right\rangle$ for a fixed size and shape of the breast.
A sufficiently strong spatial temperature variability is well distinguished due to small-scale heterogeneity of biological tissue.
Characte\-ristic scales of temperature inhomogeneity may be smaller than size of working area of the antenna of device RTM -- 01--RES used for medical measurements  \cite{Sedankin-Chupina-Vesnin-2018, Sedankin-etal-2021Intracavity-Thermometry-Medicine}. 
This is physical basis for switching to measurement schemes with a larger number of points compared to the classical scheme based on only 22 points (See next subsection).
These brightness temperature calculations do not include any contribution from the sternum in order to emphasize the influence of the internal structure of the breast.

Separately, we note that the development of mathematical and computer models of the temperature fields dynamics in various organs/tissues has acquired particular importance in the last decade. This is due, among other things, to the development of new generations of radiothermometers \cite{Leushin-etal-2022radiothermograph-Vesnin}.

\subsection{Breast Thermometric Database }\label{subsec:SUBD}

We use data from real medical MWR examinations of the breast, which are collected by the efforts of Alexander Losev and colleagues (the initial database includes 302 patients, of whom 124 have a cancer diagnosis) \cite{Levshinskii-etal-2019politeh, Germashev-2022Cyber-Physical-System, Losev-Svetlov-2022, LosevLevshinskii-2017-Math-Phys-Comput-Simul}. They isolated artifacts by excluding incomplete patient records, measurements at too high or low temperature $T_{air}$, etc. 
Such datasets include measured temperatures (MWR+IR), age, ambient temperature, breast size, first day of the menstrual cycle and its duration, body type, diagnosis.
Preprocessing reduced the sample size to $M^{(real)}=196$ patients (86 diagnosed with ``cancer''), which we call the ``REAL'' dataset. 
Thus, the sample contains only the most reliable data.
Our statistical analysis does not show differences between temperature distributions in the left and right breasts.

\begin{figure}[h!]
\begin{center}
    \includegraphics[width=0.82\hsize]{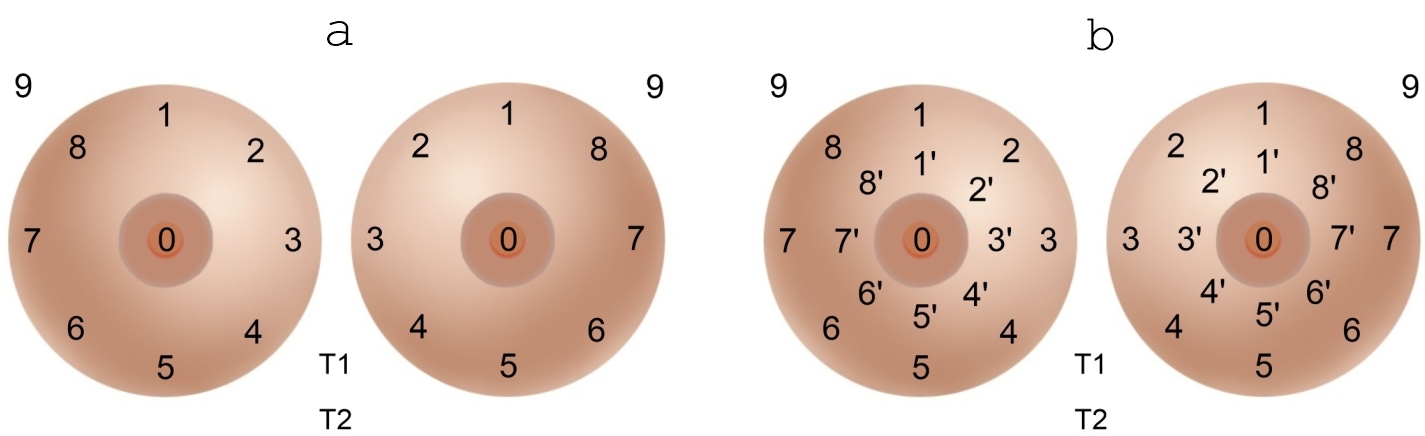}
\end{center}
\caption{\label{fig:Scheme_MG} 
The scheme for measuring the temperature of the breast according to the standard method of MWR + IR examinations contains: 9 points on the surface of each breast ($0, 1, 2,\ldots , 8$), axillary point in the area of the lymph nodes (9), two points at the sternum bottom (T1 and T2), {which gives 22 points in total} (a). The extended set of antenna location points on the breast surface was proposed and studied by us in this paper, that contains 38 points~(b).
}\end{figure}

The data of breast temperature measurements in patients were obtained in accordance with the traditionally used scheme based on 22 points (Figure \ref{fig:Scheme_MG}\,a) \cite{Polyakov-etal-2021ourVest, Losev-etal-2022Machine-Learning, Levshinskii-etal-2020-application-data-mining}.
Figure~\ref{fig:Scheme_MG}\,b also shows the new measurement scheme at 38 points on the patient's body. The effectiveness of this scheme is analyzed below in Section~3.
When considering only one breast, the traditional scheme is 9-point and the modified scheme is 17-point.

The large proportion of ``cancer'' diagnoses in the ``REAL'' sample makes it unrepresentative of testing tasks aimed at diagnosing cancer.
However, such sample has advantages for training classifiers, both machine learning algorithms and neural networks, which is our goal.

The disadvantage of the dataset ``REAL''  is its small size. In particular, cases of positive diagnoses obviously do not cover all possible variants of tumor parameters, the number of types of which is large. More importantly,  the database, for obvious reasons, does not contain any data on the actual characteristics of the tumor, such as spatial position, size, and rate of heat release.
This seems natural for medical dataset, but leads to difficulties in processing data and training neural networks. Therefore, we expand the dataset ``REAL''  using the results of numerical simulations, which form the sample ``SIMULATION''. The size of the last sample can be very large ($M^{(sim)} \gg M^{(real)}$) with known data on all tumor characteristics.

The sample structure study is based on the consideration of the hypothesis about the difference in temperatures of patients of classes ``$H$'' (Healthy) and ``$S$'' (Sick) through the use of the unsupervised learning artificial neural networks.
The initial data clustering is based on the Kohonen self-organizing maps \cite{Kohonen-lit}. A feature of this method is the reduction of the topological neighborhood during the learning process in order to ensure convergence to nearest neighbors.

\begin{figure}[h!]
\begin{center}
    \includegraphics[width=0.95\hsize]{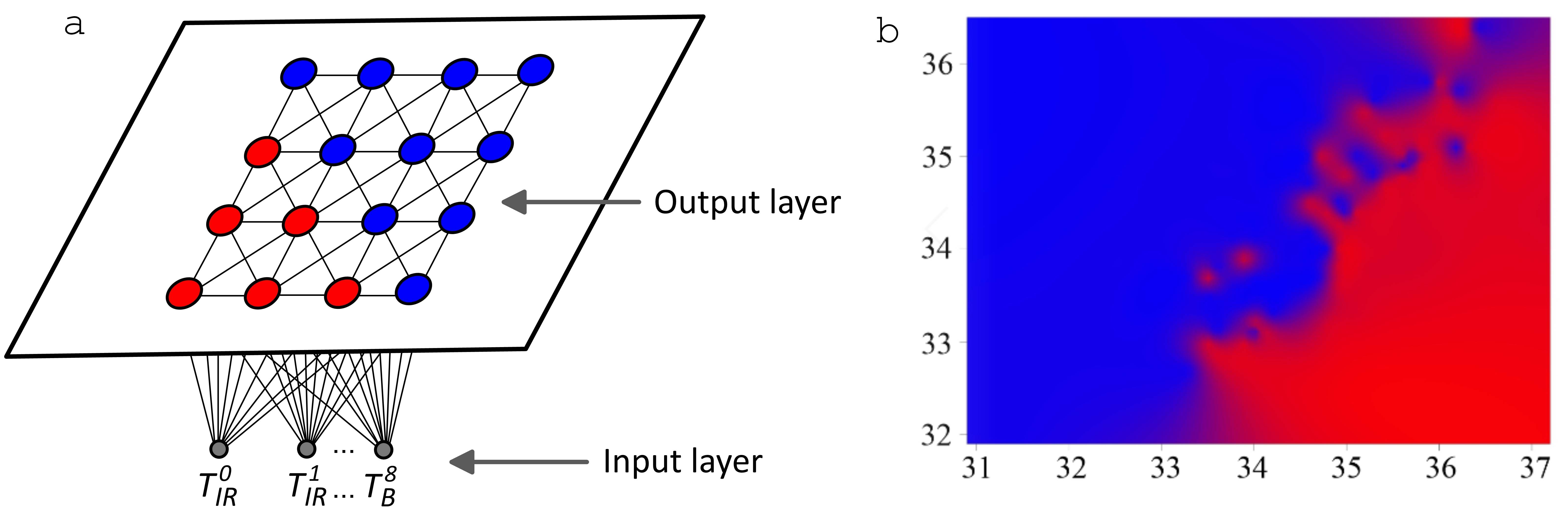}
\end{center}
\caption{\label{fig:kohonnen_map} 
Clustering scheme using the Kohonen's self-organizing map (a). An example of Kohonen map in the projection on the temperature plane: the class of models ``Sick'' is marked in red, the blue color highlights the class of models ``Healthy'' (b) (the horizontal axis is the temperature ($^\circ$C) at the point ``0'', the vertical axis is the temperature ($^\circ$C) at point ``3'').  
}\end{figure}

The input vector contains 18 temperature values $\vec{T} = (T_{IR}^0, T_{IR}^1, \ldots, T_{IR}^8, T_B^1, \ldots, T_{B}^8)$ (Figure \ref{fig:kohonnen_map}\,a).
Output maps reflect the dependencies between temperature data for different model classes (Figure \ref{fig:kohonnen_map}\,b).
The result of cluster analysis is to identify the dependence of temperature distribution on the presence of a tumor in the model. The data is divided into 2 classes, which have a characteristic structure. This, in turn, confirms the presence of consistency in the original sample.

\subsection{Validation of Computer Model for Diagnostics
Oncological Diseases Based on Machine Learning}

\begin{figure}[h!]
\begin{center}
    \includegraphics[width=0.8\hsize]{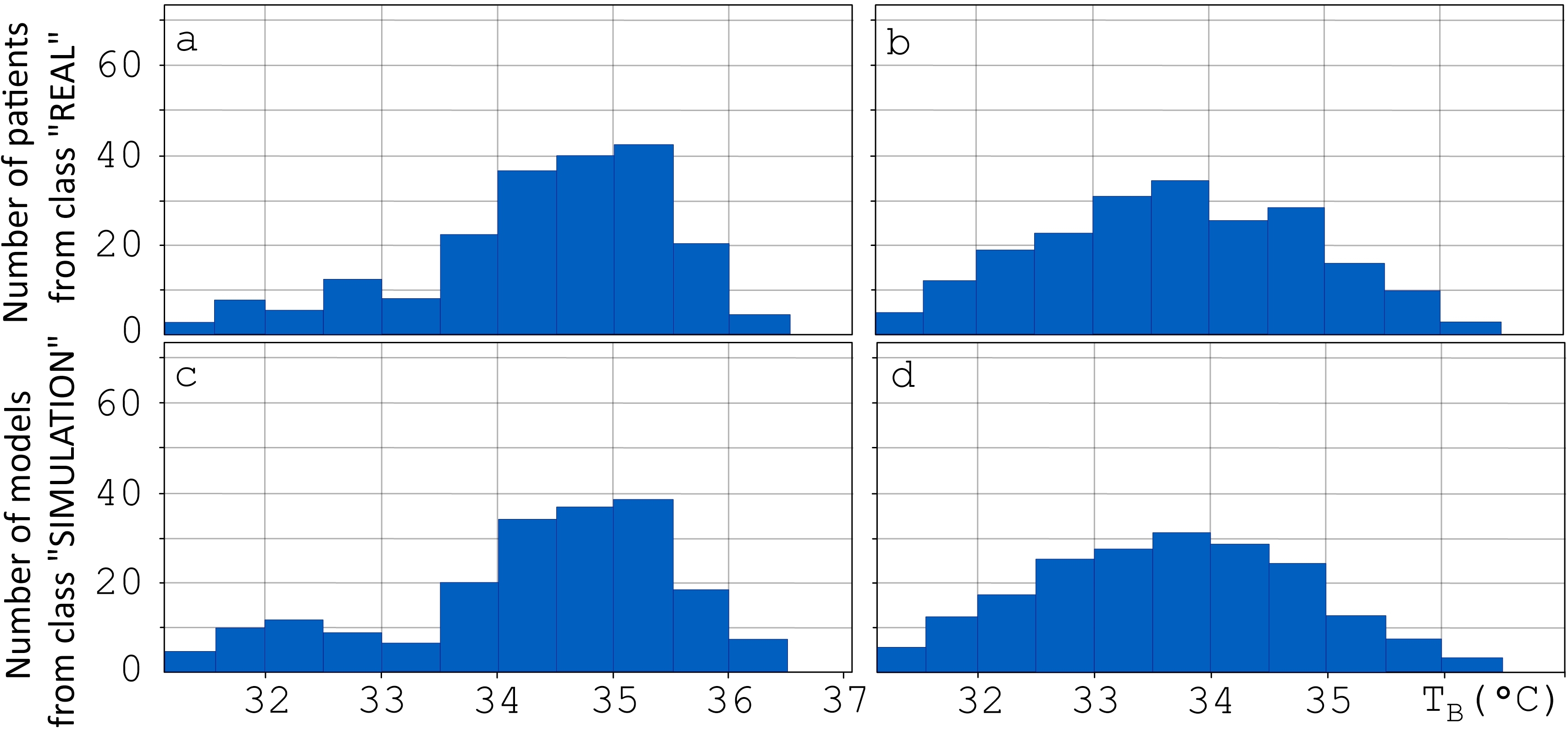}
\end{center}
\caption{\label{fig:stat_dens} 
Distributions of internal temperature values at point ``0'' according to the following classes: ``Healthy'' according to medical examinations (``REAL'') (a); dataset ``REAL'' if diagnosed with cancer~(b); ``Healthy'' according to the results of computational experiments (dataset ``SIMULATION'')~(c); ``Cancer'' by dataset ``SIMULATION''~(d).
}\end{figure}

\begin{figure}[h!]
\begin{center}
    \includegraphics[width=0.99\hsize]{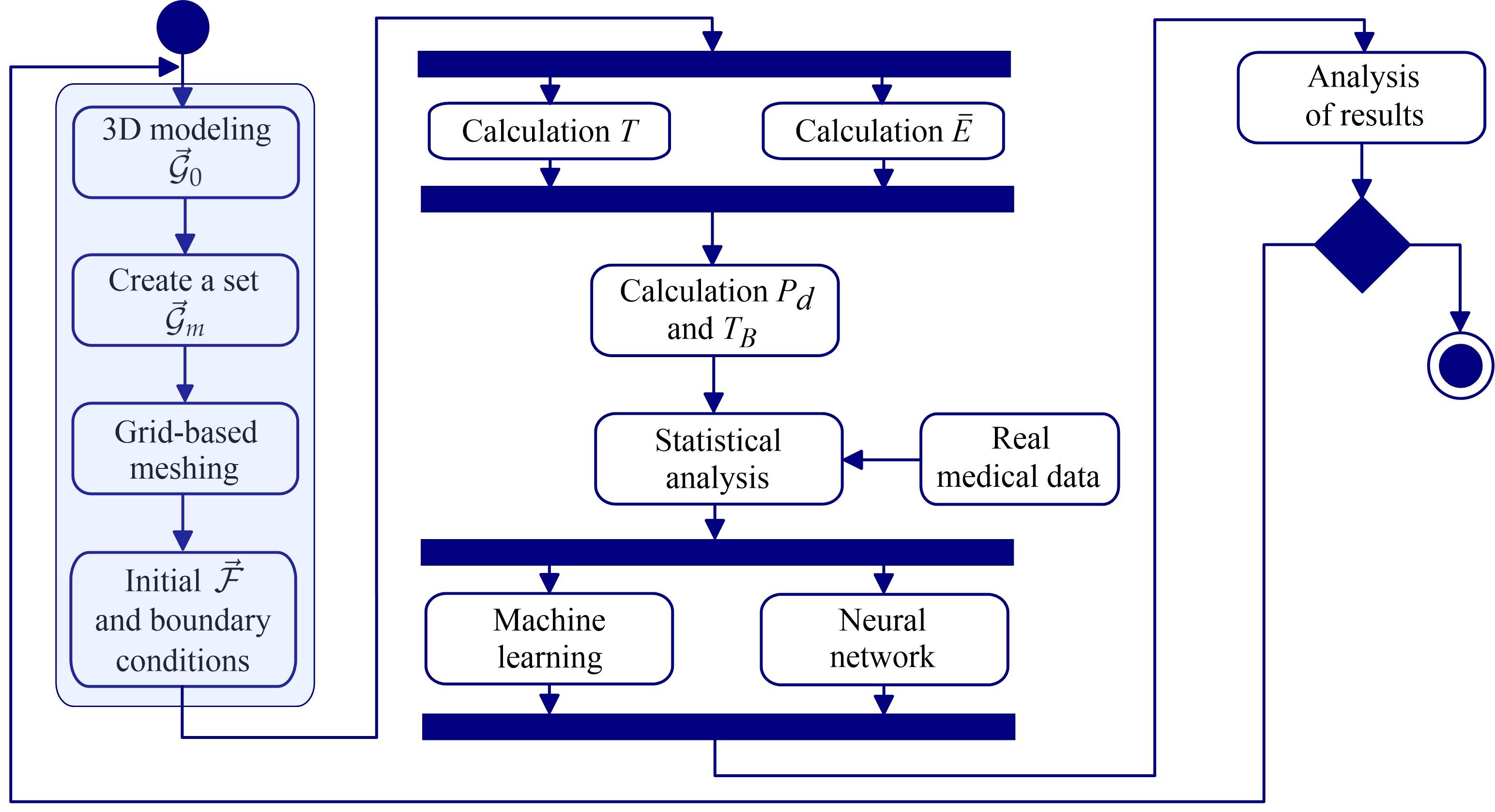}
\end{center}
\caption{\label{fig:diagram-3D-reconstruction} 
Algorithm for conducting a series of computational experiments for  large sample of breast models and processing temperature data using artificial intelligence methods.
}\end{figure}

We apply an iterative validation method to improve the quality of simulation models (Figure~\ref{fig:diagram-3D-reconstruction}). 
A distinctive feature of our approach is the ability to obtain recommendations for changing the geometry of the internal structure ($\vec{\cal G}$) and the physical characteristics of biological components ($\vec{\cal F}$) for computational experiments. This allows us to bring the simulation results and data of medical measurements into statistical agreement with each other. The proposed method is universal and applicable to a wide range of tasks, since it is not tied to the structure of the input or output data.

Validation of simulation results is provided by the following algorithm:
\begin{enumerate}
\item Building a classifier SVM over a slice with the dataset ``REAL''  for the class ``$H$'' and classifying the dataset ``SIMULATION''.
\item Building a classifier SVM over a slice with the dataset ``SIMULATION'' for the class ``$H$'' and classifying the dataset ``REAL''.
\item Analysis of the results and values of characteristics that the classifier considers incorrect.
\item The model parameters are changed for the subsequent generation of new data as a result of a series of simulations with the return to step 1 if necessary.	
\end{enumerate}

The statistical properties of the ``REAL'' and ``SIMULATION'' samples must be matched using the following procedure.
We do not distinguish between the right and left breasts and construct the distribution functions of the temperature measurements for each of the 9~points so that the ``REAL'' and ``SIMULATION'' samples are close.
Figure \ref{fig:stat_dens} shows examples of such distributions for point ``0'' (See Figure \ref{fig:Scheme_MG}). 
This figure demonstrates the statistical estimates of comparison of the final sample based on the results of numerical simulations with the data from real medical measurements for two classes of patients (the healthy patients and the tumor presence). These distributions allow us to state that the classes of model and real patients correlate well with each other.
 
\begin{figure}[h!]
\begin{center}
    \includegraphics[width=0.995\hsize]{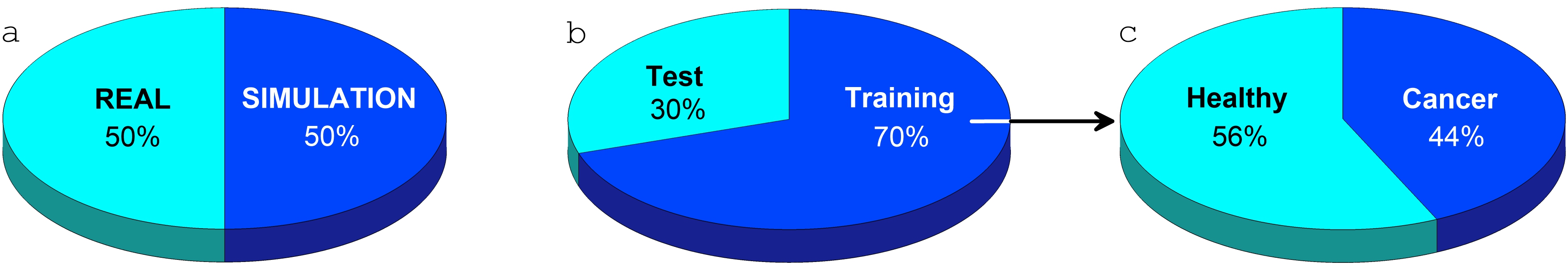}
\end{center}
\caption{\label{fig:set_structure} 
Sample structure used in machine learning algorithms
}\end{figure}

The dataset for machine learning contains 392 units (196 units in the ``SIMULATION'' sample and 196 units in the ``REAL'' sample) (Figure \ref{fig:set_structure}a), of which 172 units have a cancer diagnosis. The training set contains 274 objects and the test set size is 118 (Figure \ref{fig:set_structure}b). To correctly assess the performance of the algorithms, a cross-validation procedure is carried out. The training sample contains 153 units of the ``Healthy'' class and 121 units of the ``Cancer'' class (Figure \ref{fig:set_structure}c).

\subsection{Artificial Intelligence Algorithms for Temperature Data Processing}

\begin{figure}[h!]
\begin{center}
    \includegraphics[width=0.998\hsize]{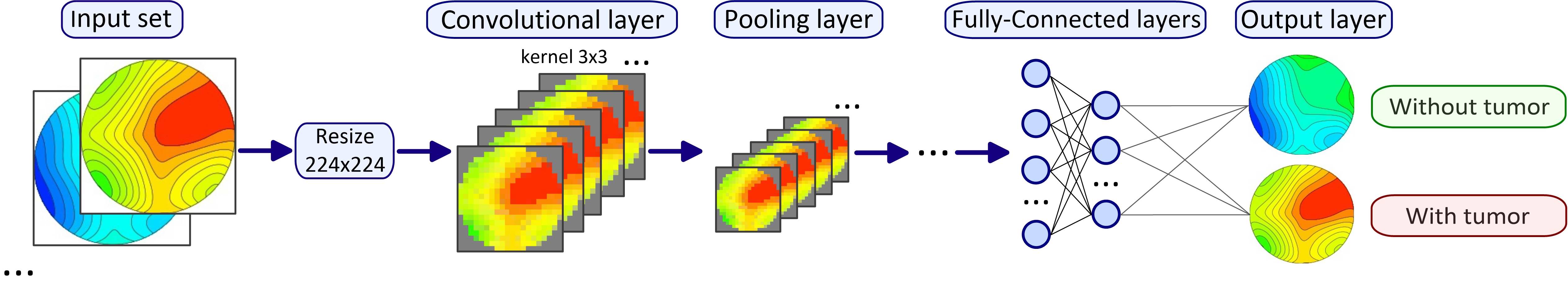}
\end{center}
\caption{\label{fig:CNN_structure} 
Used CNN scheme for binary classification of thermometric data.
}\end{figure}

Artificial neural networks are a universal information processing tool and are widely used in various applications of medical diagnostics \cite{Ekici-Jawzal-2020Breast-cancer-neural-networks, Abdel-Jaber-2022, LoandYin-2021}. We use the Convolutional Neural Network (CNN) based on the VGG16  architecture \cite{SimonyanZisserman}. Figure \ref{fig:CNN_structure} demonstrates the scheme of artificial neural network for the problem of binary classification of MWR data.

Our software module implements the random partition method to form sets of images, on which the neural network is subsequently trained and tested.
The method includes the creation of sets by random selection of data. Sensitivity, specificity and classification efficiency are calculated as the arithmetic mean of all partitioning results.
It is important to control the number of epochs in the process of training the neural network, so that there is no overfitting regime.

NVIDIA CUDA parallel computing technology was used to increase the speed of the neural network. We performed neural network classification using the QUADRO RTX 4000 graphics card based on the NVIDIA Turing architecture (Nvidia Corporation).

We use the following machine learning algorithms for binary classification of temperature data: support vector machine (SVM) \cite{SVMChen}, k nearest neighbor method (KNN) \cite{Sutton} and naive Bayes classifier (NBC) \cite{Berrar}.
The Gaussian kernel with a radial basis function and parameter $\gamma = 0.7$ underlies the SVM algorithm. We apply the weighted voting method for the KNN algorithm with the parameter $k = 5$.
The effectiveness measure of medical diagnostics is the geometric mean of sensitivity and specificity
\begin{equation}\label{eq-Sens-Spec}
	eff = \sqrt{sens \cdot spec} \,,
\end{equation}
where $\displaystyle sens = \frac{TP}{TP + FN}$, $\displaystyle spec = \frac{TN}{TN + FP}$, $TP$ is the number of correctly classified breasts for class ``Cancer'', $FN$ is the number of misclassified breasts for class ``Cancer'', $TN$ is the number of correctly classified breasts for class ``Healthy'', $FP$ is the number of breasts that are misclassified classified as ``Healthy''. 

The F1-score value is an integral estimate of the precision and recall of classifiers, which calculates the harmonic mean using the formula
\begin{equation}\label{eq-F1_score}
F1 = \frac{2\cdot precision\cdot recall}{precision+recall}\,,
\end{equation}
where $\displaystyle precision=\frac{TP}{TP+FP}$ and $\displaystyle recall=\frac{TP}{TP+FN}$.

Additionally, we calculate the Matthews correlation coefficient (or Phi coefficient) for all methods
\begin{equation}\label{eq-Matthews}
\phi = 	 \frac{TP\cdot TN-FP\cdot FN}{\sqrt{(TP+FP)(TP+FN)(TN+FP)(TN+FN)}}\,.
\end{equation}
The Phi coefficient is a measure of the quality of binary classifications in the case of markedly different samples in size, since the number of patients is small compared to the number of healthy people in real medical screening conditions.  

We mix the data in the test set many times to correctly assess the classification efficiency. This provides the method of randomly splitting the sample into 5 overlapping subsamples in which the ratio of Healthy to Sick is the same and corresponds to the parent population. Thus, the representativeness of the sample is preserved. Classification results are averaged over 5 blocks.

\section{Results}\label{sec:result}

\subsection{Conditions for Detecting Weak Tumors}

Growth of tumors at the stage of exceeding its size by several millimeters is possible only if a small capillary network with increased blood flow is formed around it \cite{Badiger-2020, Yahara-2003tumor-growth-due-to-blood-flow, Morales-Guadarrama-2021}.
Therefore, the tumor focus can be quite a powerful source of heat in the breast. Malignant neoplasms often have extremely high heat release rates relative to other biological components, especially in the early stages of disease development (See Figure \ref{fig:cancer-energy}).

\begin{figure}[h!]
\begin{center}
    \includegraphics[width=0.95\hsize]{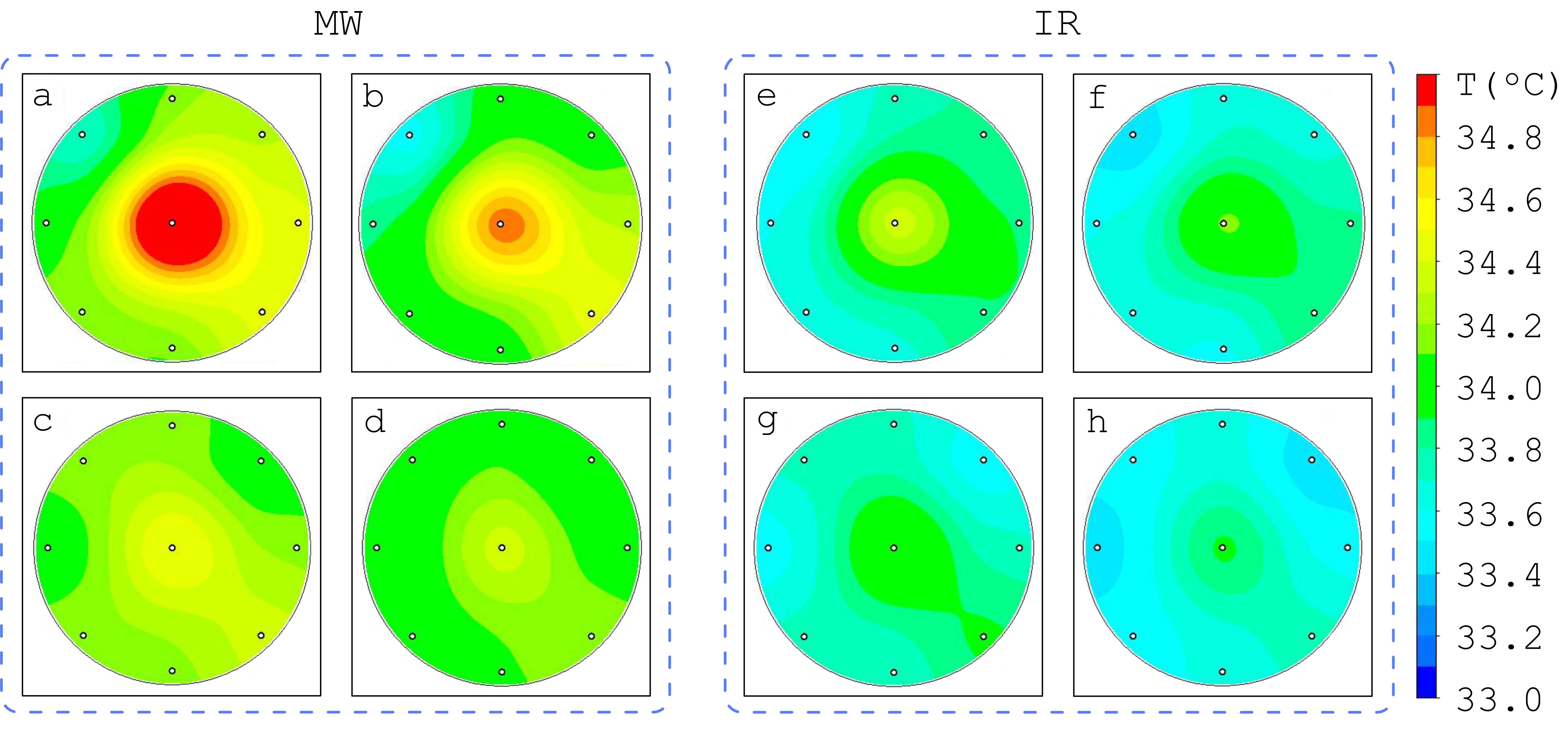}
\end{center}
\caption{\label{fig:Temp}  
Brightness and infrared temperature distributions of three models with different tumor sizes: a, e) $R=1$ {cm}; b, f)~$R=0.75$ {cm}; c, g) $R=0.5$ {cm}. The models in panels d, h) does not contain tumor.
}\end{figure}

We considered separately the problem of detecting weak tumors, which are characteristic of the initial stage of the disease and are poorly distinguished by traditional methods such as mammography. Special datasets were constructed for simulations containing tumors of different radii ($R$ = 1 cm, $R$ = 0.75 cm, $R$ = 0.5 cm) at $Q^{(can)}=3\cdot 10^4 $\,W\,m$^{-3}$. Figure~\ref{fig:Temp} shows the effect of tumor size in models with the same internal structure of the breast, when only the value of $R$ changes and the tumor is located on the axis of symmetry passing through the nipple. Even such simple direct images indicate the difficulty of detecting weak tumors. IR data on the temperature of the skin surface gives a poorer opportunity to detect tumors (See Figure~\ref{fig:Temp}).
 
\begin{figure}[h!]
\begin{center}
    \includegraphics[width=0.68\hsize]{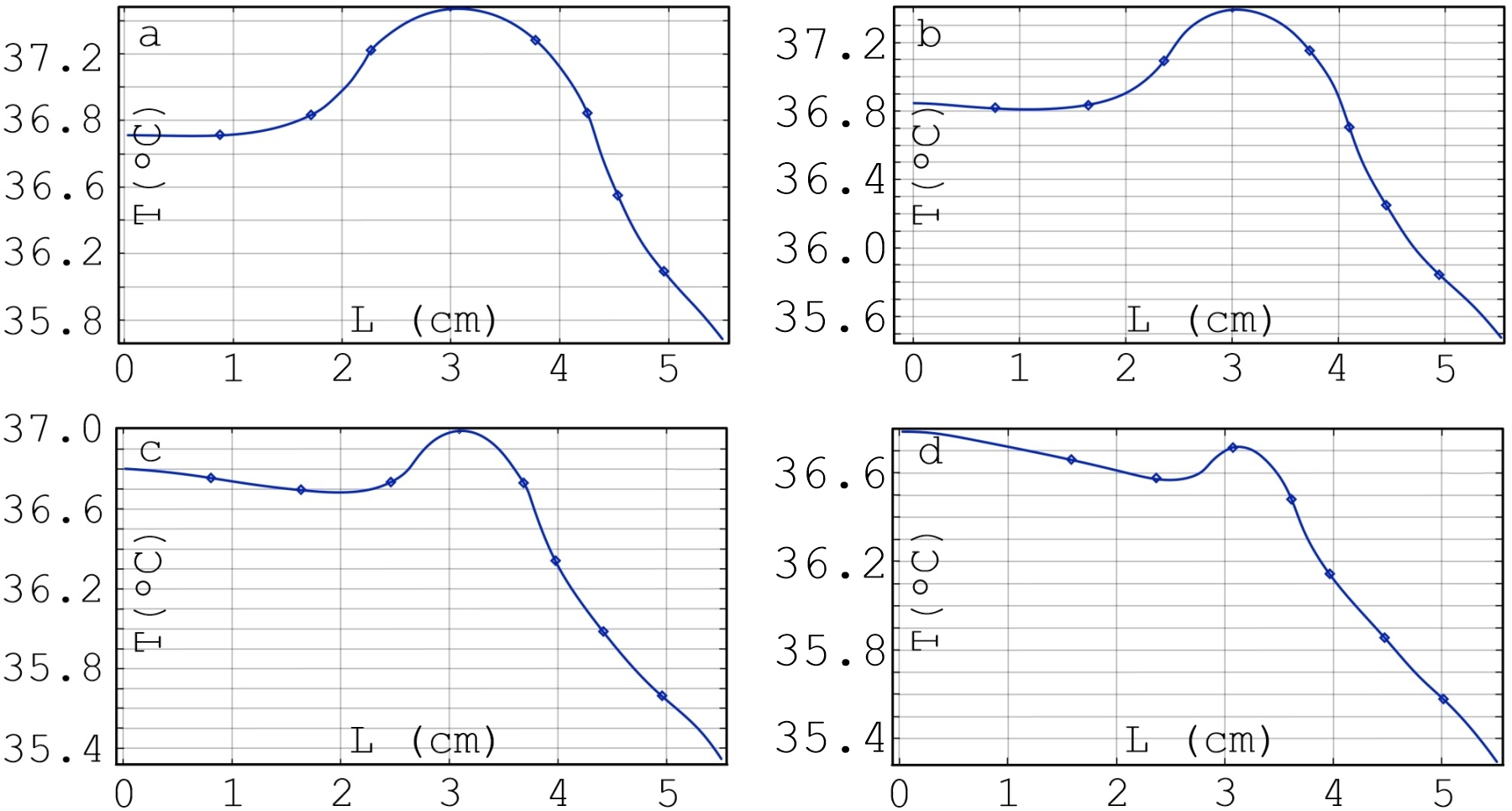}
\end{center}
\caption{\label{fig:radius_profile} 
Distributions of thermodynamic temperature over depth for tumors of different radii $R$ at $Q^{(can)} = 3\cdot 10^4 $\,W\,m$^{-3}$: a) $R$ = 1 cm, b) $R$ = 0.75 cm, c) $R$ = 0.5 cm, d) $R$ = 0.35 cm.
}\end{figure}

\begin{figure}[h!]
\begin{center}
    \includegraphics[width=0.6\hsize]{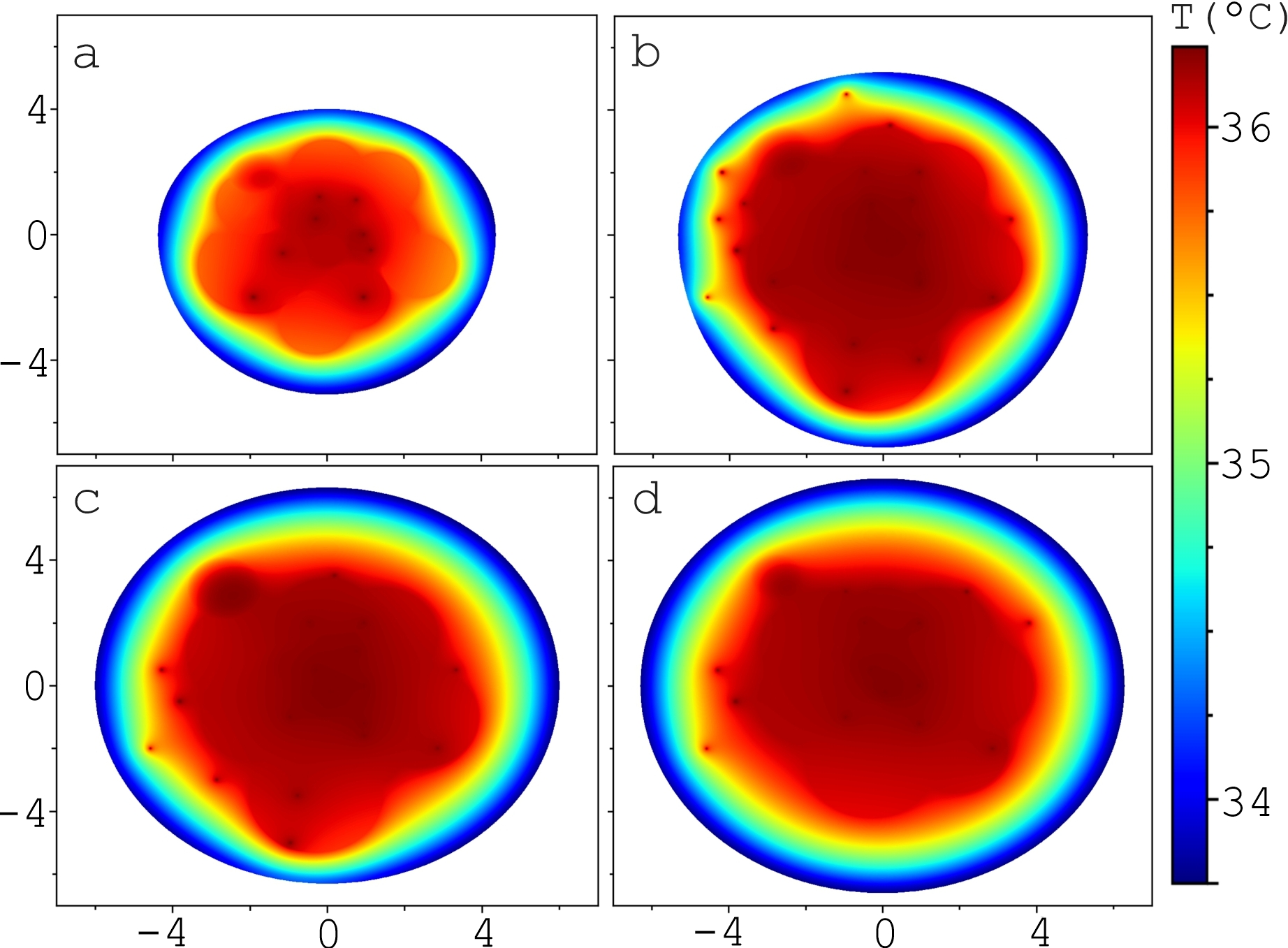}
\end{center}
\caption{\label{fig:temperature_z}
Thermodynamic temperature distributions at different depths from the nipple on flat sections: $L^{(can)}=2$ cm (a); $L^{(can)}=3$ cm (b); $L^{(can)}=4$ cm (c); $L^{(can)}=5$ cm (d). The tumor is in the vicinity of the measurement point ``8''.
}\end{figure}

The brightness temperature $T_B$ is not a local characteristic, since it is calculated as an integral of some volume (See (\ref{eq-TBend})). The thermodynamic temperature $T$ is more sensitive to small-scale tissue inhomogeneities and heat sources. Figures \ref{fig:radius_profile} and \ref{fig:temperature_z} show temperature distributions ($T$) from the nipple to the sternum in different models, which clearly indicate the tumor presence even at $R=0.35$ {cm}, but the contribution of such local formations to the brightness temperature is small even at $ Q^{(can)}=3\cdot 10^4 $\,W\,m$^{-3}$. Colder small tumors at $Q^{(can)} < 2\cdot 10^4 $\,W\,m$^{-3}$ are not detected at all, since the integral heat power from the tumor is proportional to $Q^{(can)}\cdot R^3$.

\begin{figure}[h!]
\begin{center}
    \includegraphics[width=0.65\hsize]{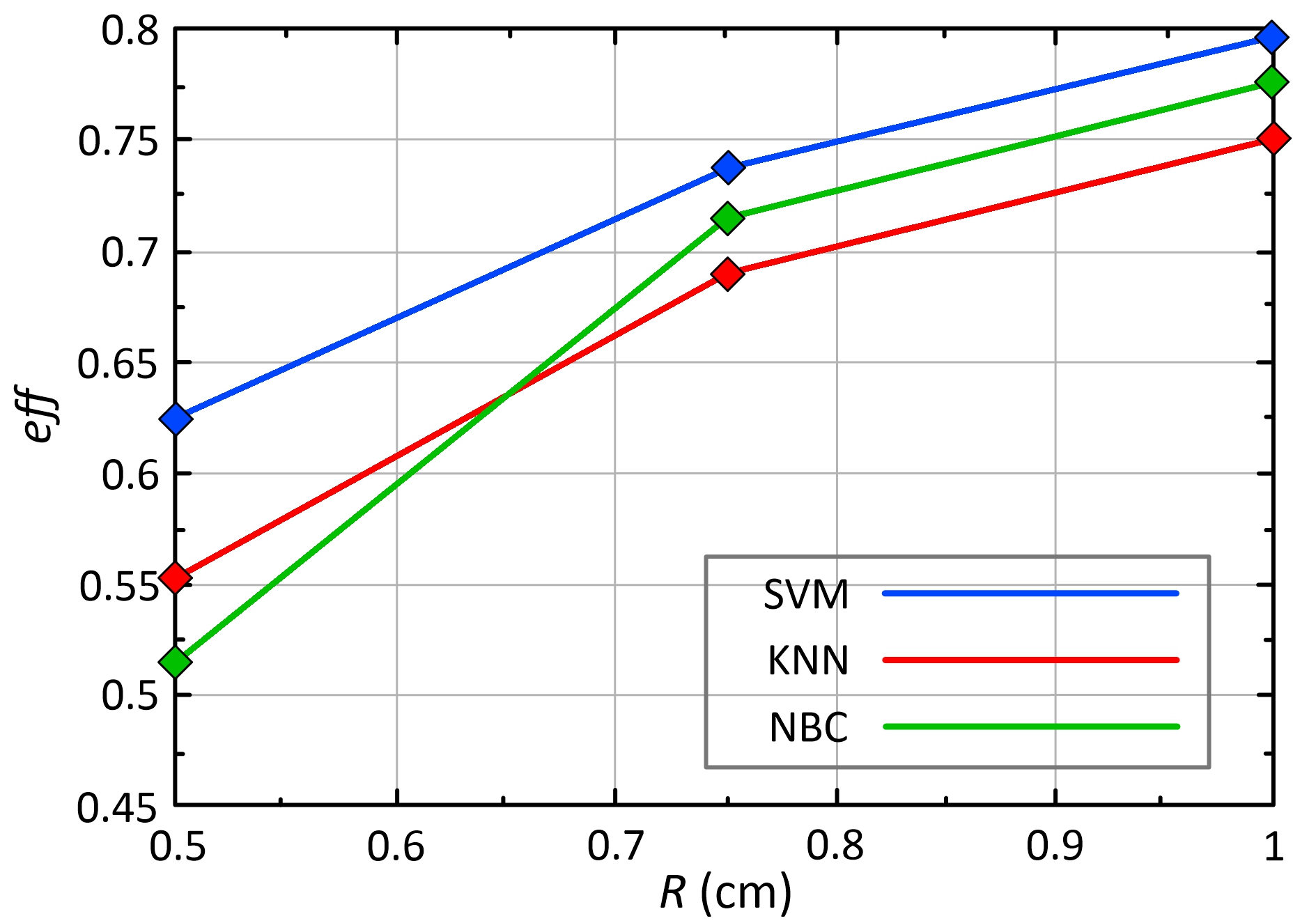}
\end{center}
\caption{\label{fig:radius_dependency} 
The effectiveness of MWR diagnostics ($eff$) vs the tumor radius for various machine learning methods.
}\end{figure} 

\subsection{Influence of Tumor Spatial Location on the Brightness Temperature}

The brightness temperature $T_B$ is an integral value and depends both on the thermodynamic temperature $T$ and on the distribution of the electric field strength $\vec{E}$ in the biological tissue.
The factor of the spatial location of the tumor inside the breast significantly affects the results of temperature measurements and diagnosis.
If two models ($a$ and $b$) contain tumors at different depths $L^{(can)}$, then the deviations of the brightness temperature values in these models can be conveniently characterized by the formula
\begin{equation}\label{slae}
	\psi_{ab}=\frac{1}{N}\sum_{i=0}^{N=8} \frac{|T_{i}^{a}-T_{i}^{b}|}{T_{i}^{b}},
\end{equation}
where the $i$-th index denotes the position of the antenna on the surface of the breast according to the measurement technique (See Figure \ref{fig:Scheme_MG}).

Table \ref{tumorlocation} contains the results of calculations using the formula (\ref{slae}), which shows that the maximum deviations in the brightness temperature distribution can reach almost 6~percent, which is significant for medical diagnosis.

\begin{table}[h!]
\caption{
Relative brightness temperature deviations $\psi_{ab}$ for two models with different tumor localizations in depth ($L^{(can)}$).
}\label{tumorlocation}
\begin{center}
\begin{tabular}{cccccc}
\hline
$L^{can}$ & {1 cm} & {2 cm} & {3 cm} & {4 cm} & {5 cm} \\ \hline
{1 cm}  & 0     & 0.016 & 0.027 & 0.041 & 0.059 \\
{2 cm}  & 0.016 & 0     & 0.011 & 0.026 & 0.043 \\
{3 cm}  & 0.027 & 0.011 & 0     & 0.015 & 0.033 \\
{4 cm}  & 0.041 & 0.026 & 0.015 & 0     & 0.019 \\
{5 cm}  & 0.059 & 0.043 & 0.033 & 0.019 & 0     \\ \hline
\end{tabular}
\end{center}
\end{table} 

Figure \ref{fig:radius_dependency} shows the results of calculating the efficiency (\ref{eq-Sens-Spec}) of three machine learning methods for diagnosing the presence of a tumor. 
The dataset includes the results of modeling brightness and IR temperatures according to the classical measurement scheme (See Figure \ref{fig:Scheme_MG}\,a).
The efficiency exceeds 0.75 for radius of 1 cm and above, and the scatter for all three algorithms is within 5 percent. Even small tumors with radius of $R=0.5$ cm allow the correct definition of the class ``Cancer'' with probability of 62.5 percent. Support Vector Machine (SVM) gives a better result compared to other machine learning methods (KNN, NBC), which indicates that this method is better suited for this type of problem and for this training dataset structure. The gain of the SVM method with respect to NBC is 10~percent for dataset with tumor $R = 0.5$~cm, which is significant for the problem of medical diagnostics. The efficiency of the k nearest neighbor method becomes unacceptable for small tumor radii ($R\le 0.5$~cm).

\begin{figure}[h!]
\begin{center}
    \includegraphics[width=0.5\hsize]{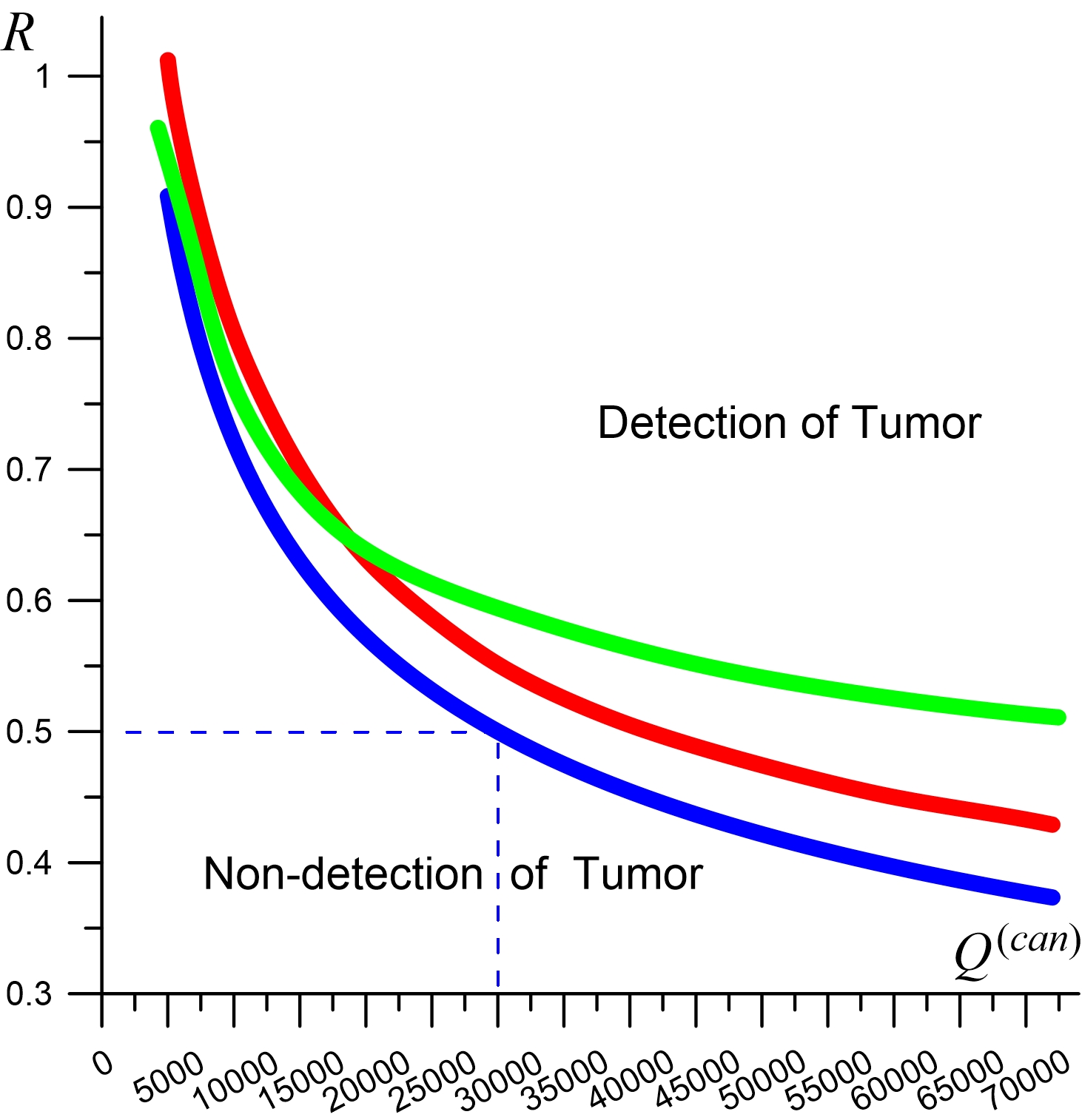}
\end{center}
\caption{\label{fig:Qcar-R-crit} 
Tumor detection boundaries on the plane of parameters $R$ ([cm]) and $Q^{(can)}$ ([W$\cdot$m$^{-3}$]) by different machine learning methods (blue line is SVM , green line is NBC, red line is KNN).
}\end{figure} 

Figure \ref{fig:Qcar-R-crit} demonstrates the integral result of evaluating the possibility of detecting a tumor depending on the size and specific energy release, which is determined by the tumor doubling time. Thick lines of different colors show the boundaries separating the parameter areas at which one or another method can detect a tumor. Heat release rates of $Q^{(can)} = 30\,000$\,W\,m$^{-3}$, which is typical of fast growing tumors doubling in 100 days or less, allow detection of tumors up to 1 cm in diameter.
The classification of even smaller tumors tumors requires a transition to the analysis of feature spaces, an increase in the sample size, and the use of heuristics \cite{Losev-etal-2022Machine-Learning,  LosevLevshinskii-2017-Math-Phys-Comput-Simul, Levshinskii-etal-2020-application-data-mining}. Feature spaces are built in the form of matrices of various dimensions, the elements of which are temperature differences at different measurement points (See Figure \ref{fig:Scheme_MG}). Moreover, both brightness temperature and infrared temperature are used.

We note the study \cite{Bardati-2008}, in which a tumor with a radius of {0.5 cm} is effectively detected by the MWR method when it is located at a depth of no more than 2.8 cm, which is consistent with our results. However, the authors use a multi-layer tissue model, which makes it easier to find the hot spot against the background of an almost uniform temperature distribution. Moreover, the tumor sign is a local increase in temperature by 0.1 degrees, but such fluctuations are also natural in the absence of a tumor (See Figure \ref{fig:bright-temperature-8fig}).

\begin{figure}[h!]
\begin{center}
  \includegraphics[width=0.68\hsize]{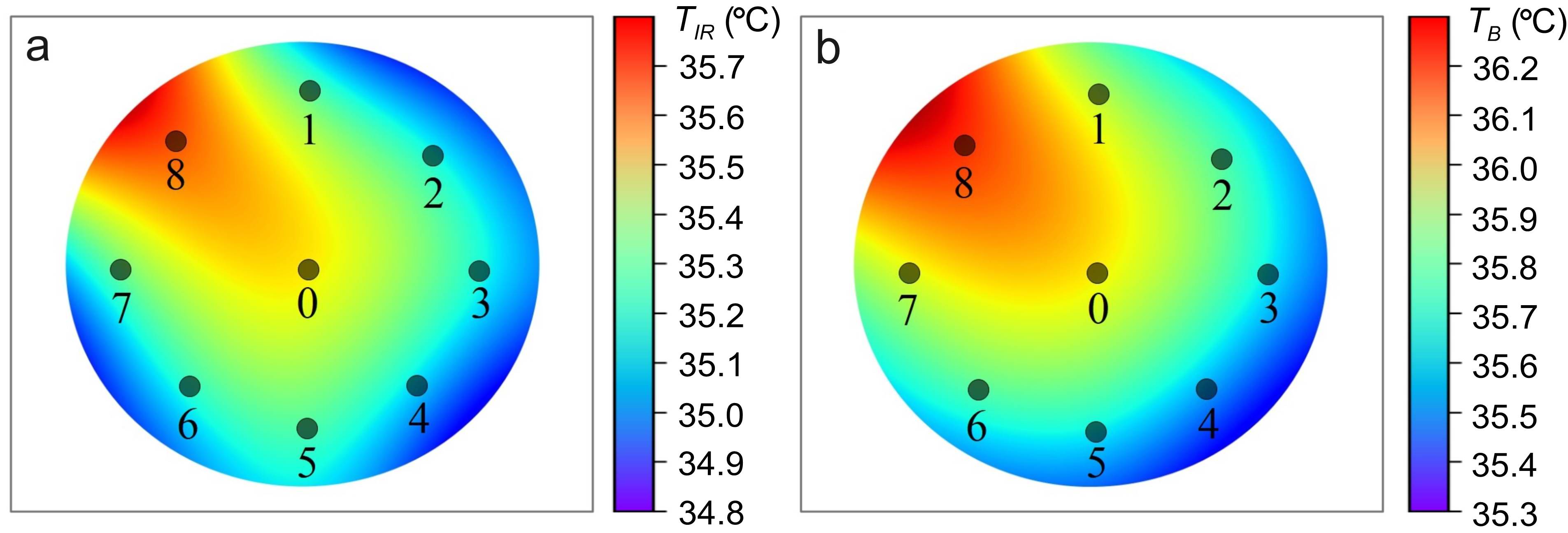}
\end{center}
\caption{
The brightness and infrared temperature distributions from the simulations show a characteristic bell-shaped appearance.
 \label{fig:bright-temperature}}
\end{figure}

A feature of the high specific heat release of the tumor is the formation of the so-called ``hot’’ bell shape (Figure~\ref{fig:bright-temperature}), which is an additional dimension in the feature space for CNN.

\subsection{Application of Artificial Neural Networks for MWR Data }

Let's expand the combined dataset (``REAL''+``SIMULATION'') from Subsection 3.2 with additional numerical simulation results by varying the size and location of the tumor.
Table~\ref{tab:CNN} contains results of binary classification of brightness temperature and IR temperature for four different neural network topologies. A significant influence of the structure of the neural network classifier on the efficiency of diagnostics is shown. The difference between the best (Topology 3) and worst (Topology 2) efficiency values reaches 21 percent. At the same time, a larger number of internal layers (Topology 1) does not give an advantage in the final results.

\begin{table}[h!]
\caption{
Binary classification results for artificial neural networks with different fully-connected layer architecture.
}\label{tab:CNN}
\begin{center}
\begin{tabular}{lcccc}
\hline
                                   & Topology 1 & Topology 2 & Topology 3 & Topology 4 \\ \hline
Number of layers                   & 8       & 6       & 5       & 4       \\
Number of neurons on the 1st layer & 20      & 20      & 20      & 20      \\
Number of neurons on the 2nd layer & 20      & 20      & 20      & 10      \\
Number of neurons on the 3rd layer & 20      & 10      & 14      & 3       \\
Number of neurons on the 4th layer & 20      & 6       & 3       & 2       \\
Number of neurons on the 5th layer & 20      & 4       & 2       & --       \\
Number of neurons on the 6th layer & 20      & 2       & --       & --       \\
Number of neurons on the 7th layer & 20      & --       & --       & --       \\
Number of neurons on the 8th layer & 2       & --       & --       & --       \\
$sens$                               & 0.74    & 0.66    & 0.86    & 0.81    \\
$spec$                               & 0.67    & 0.61    & 0.82    & 0.71    \\
$eff$                               & 0.7     & 0.63    & 0.84    & 0.76    \\
 F1 & 0.72 & 0.61 & 0.83 & 0.75 \\ 
$\phi$ & 0.46 & 0.27 & 0.63 & 0.54  \\ \hline
\end{tabular}
\end{center}
\end{table}

\begin{table}[h!]
\caption{Confusion matrix for CNN Topology 3}\label{tab:Confusion-matrix}
\begin{center}
\begin{tabular}{cccc}
\hline
\multicolumn{2}{l}{}                                                                                                                                                        & \multicolumn{2}{c}{Predicted condition}                                                                       \\ \hline
                                                                                                 & \multicolumn{1}{c|}{\begin{tabular}[c]{@{}c@{}}Total\\ 118\end{tabular}} & \begin{tabular}[c]{@{}c@{}}Positive\\ 56\end{tabular} & \begin{tabular}[c]{@{}c@{}}Negative\\ 62\end{tabular} \\ \cline{1-2}
\multicolumn{1}{c}{{\begin{tabular}[c]{@{}c@{}} \\ Actual \\ condition\end{tabular}}} & \begin{tabular}[c]{@{}c@{}}Positive\\ 51\end{tabular}                    & \begin{tabular}[c]{@{}c@{}} \\ 44\end{tabular}        & \begin{tabular}[c]{@{}c@{}} \\ 7\end{tabular}         \\
\multicolumn{1}{c}{}                                                                             & \begin{tabular}[c]{@{}c@{}}Negative\\ 67\end{tabular}                    & \begin{tabular}[c]{@{}c@{}} \\ 12\end{tabular}        & \begin{tabular}[c]{@{}c@{}} \\ 55\end{tabular}        \\ \hline
\end{tabular}
\end{center}
\end{table}

The value of the mean harmonic measure F1 is close to the value of $eff$ within about 3 percent (See Table 3), which indicates correct classification.
The $\phi$ coefficient (also called the Matthews correlation coefficient (MCC)) turns out to be noticeably smaller than the $eff$ and F1-score values, but also indicates the advantage of Topology 3 and the disadvantages of Topology 2. The best value of $\phi = 63$ percent for Topology 3 in our case is significantly less than for complex diagnostic methods, such as ultrasound elastography \cite{Mao-etal-2022Elastography}, mammography \cite{Ragab-etal-2021breast-cancer-Multi-DCNNs-mammogram}, for which F-score and MCC $\geq 90$ percent is a typical result.
This difference is due to the use of fundamentally more accurate physical methods that make it possible to directly visualize the internal three-dimensional tissue structure with a resolution of $\leq$ 0.5 cm for ultrasound \cite{Akatsuka-etal-2022ultrasound} and $< 1$ cm for mammography \cite{Rangarajan-et-al-2022}.
The Microwave Radiometry detects the presence of disease indirectly through temperature with a spatial resolution of approximately 2 cm in the plane and 4 cm along the normal coordinate.
An additional negative factor for MWR diagnosis is the more complex relationship between the observed temperature distributions and the presence of the disease and its stage.
Therefore, lower values of $eff$, F1, $\phi$ for MWR are a natural result. The significance of MWR diagnostics lies in the possibility of organizing mass, non-invasive and cheap cancer screening for preliminary diagnosis.

\begin{figure}[h!]
\begin{center}
    \includegraphics[width=0.9\hsize]{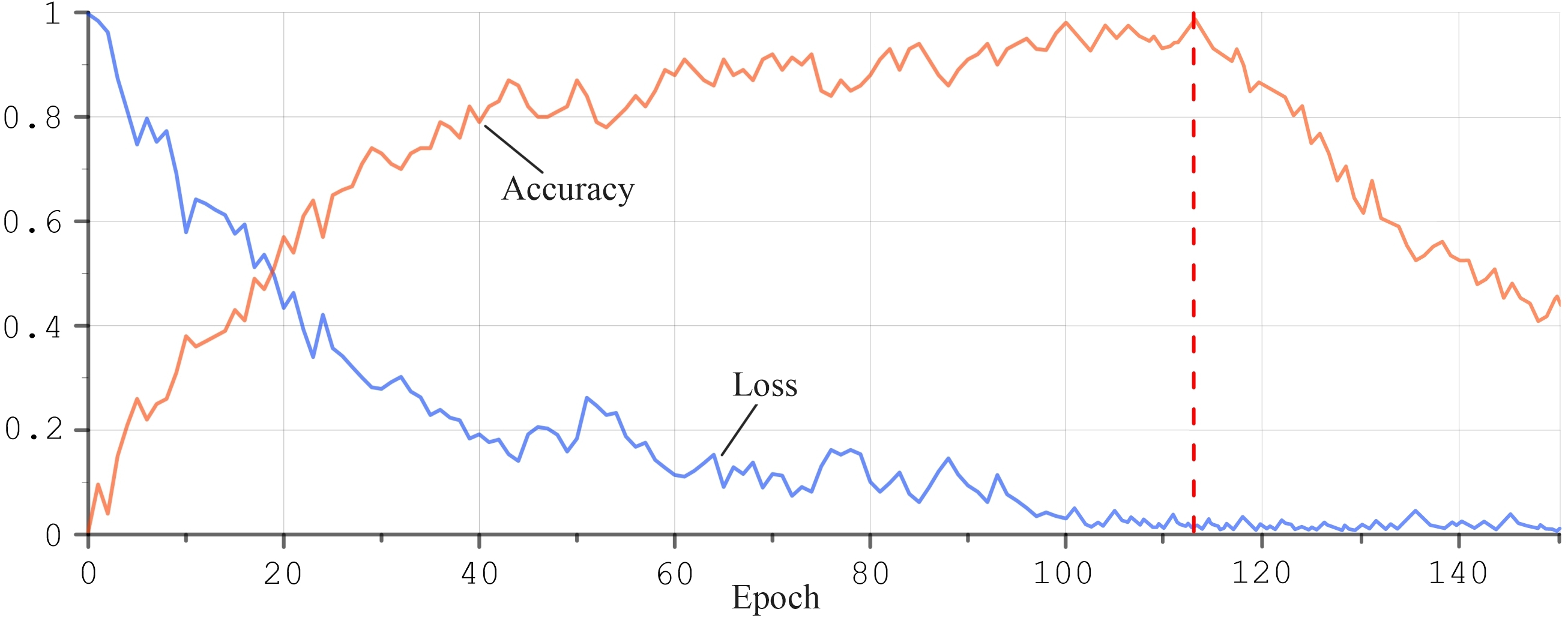}
\end{center}
\caption{\label{fig:loss_training} 
Dependences of the accuracy and the loss on the epoch. The red dashed line shows the epoch when the retraining of the neural network starts.
}\end{figure}

It is necessary to control the number of epochs in the process of training the neural network so that overfitting problems do not arise. Figure \ref{fig:loss_training} shows the moment (epoch = 113) at which the retraining of the neural network for Topology 3 occurs.
The overfitting problem is solved based on the Dropout method. This method allows regularization of artificial neural networks by eliminating random neurons in different epochs of neural network training. Additionally, we control the number of epochs in the training process.

To improve the spatial resolution of the measured IR and MW temperatures of the breasts, we propose to use the modified measurement scheme based on 17 points on one breast (See Figure \ref{fig:Scheme_MG}\,b), changing their localization and adding 8 new antenna positions to the standard measurement scheme.
A similar increase in the number of points at which brightness temperature is measured can be applied to MWR diagnostics of other organs as well (See Figure \ref{fig:MedicalUsesRTM}).

\begin{table}[h!]
\caption{
Results of binary classification of model data only using 9-point and 17-point schemes for measuring brightness temperature and infrared temperature.
 }\label{newmetodic}
 \begin{center}
\begin{tabular}{rccccc}
\hline
                  & $sens$ & $spec$ & $eff$ &  F1 & $\phi$ \\ \hline
9-point (CNN)  & 0.74 & 0.62 & 0.68 & 0.7 & 0.36 \\
17-point (CNN) & 0.79 & 0.64 & 0.71 & 0.71 & 0.44 \\
9-point (SVM)  & 0.76  & 0.69  & 0.72  & 0.71 & 0.45  \\
17-point (SVM) & 0.8  & 0.71  & 0.75  & 0.74 & 0.51 \\ 
9-point (KNN)  & 0.71  & 0.6  & 0.65  & 0.64 & 0.31  \\
17-point (KNN) & 0.75 & 0.63 & 0.69  & 0.67 & 0.38 \\
9-point (NBC)  & 0.72  & 0.62  & 0.67 & 0.65 & 0.33  \\
17-point (NBC) & 0.73  & 0.63  & 0.68  & 0.67 & 0.36 \\ \hline
\end{tabular}
\end{center}
\end{table}

The results of our binary classification in Table \ref{newmetodic} allow us to conclude that the use of an extended measurement scheme can increase the efficiency of medical diagnosis of breast cancer by 4 percent, which is a significant result. The sensitivity of such measurement scheme increases by 5 percent. Thus, such modification of the method can complement the main examination to clarify the diagnosis. The transition from 9-point scheme to 17-point scheme increases F1-score by 1--3 percent for various algorithms, which is close to $eff$. The MCC value increases by 3-8 percent for the four considered algorithms. Thus, the best classification quality is given by SVM for both 9-point scheme and 17-point scheme.

\section{Conclusions and Discussion}

Medical practice is based on a variety of methods for diagnosing breast cancer.
These are both methods for detecting abnormal structures inside the tissue (ultrasound, mammography, tomography), as well as blood tests, biomarkers.
A special place belongs to the MWR, which is based on the analysis of temperature fields due to additional heat release by the tumor.
 
Significant advances in the understanding of the clinic of breast diseases, major changes in treatment approaches and significant development of functional diagnostic methods have been made in recent decades \cite{Ekici-Jawzal-2020Breast-cancer-neural-networks, Liang-etal-2022magnetic-resonance-imaging, Hong-etal-2022Tumor-Markers, Brown-etal-2022statist, Hatwar-2017}.
However, the solution to the problem of increasing survival in breast cancer is far from acceptable level \cite{Siegel-etal-2021Cancer-Statistics-2021, Ravaioli-2020}.
The main problem is the difficulty of early detection of weak tumors.
Nearly half (30 to 50~percent) of breast cancer patients seek treatment for the first time with stage III disease, and cancer mortality is 15~percent in the United States (male mortality from breast cancer is 20 percent).
Such an unfavorable situation with mortality is developing even in the USA, although the majority of patients with breast cancer are diagnosed with an early stage of the disease~\cite{Siegel-etal-2021Cancer-Statistics-2021}.
The widespread use of ultrasound for breast cancer screening does not fundamentally solve the problem, since the method detects tumors with an average size of 1.3 cm, while {$0.5-0.7$~cm} is necessary for successful treatment.
Therefore, the transition to diagnostic methods capable of  detecting tumors smaller than 1~cm is extremely important for breast oncology.  

Our main efforts are aimed at creating hybrid methods for analyzing and extracting knowledge from thermometric data based on the joint application of machine learning algorithms and computer modeling of biophysical processes. This approach has been developed for the diagnosis of oncology in the breast. However, it seems to be quite versatile and can be applied to other organs and diseases.
Involving the results of numerical simulation of the thermal and radiation fields dynamics inside tissues with a complex structure makes it possible to use information about tumor parameters in datasets for artificial intelligence algorithms.
Such synthetic datasets, including both the real measurements and the results of numerical simulations, can improve the efficiency of medical diagnostics.

We highlight our main results below.

\noindent 1) 
We propose datasets formation method based on combining two samples. One contains the results of real temperature measurements (``REAL''). The second sample is based on simulations of thermal and radiation processes inside breast models (``SIMULATION'').
The sample ``SIMULATION'' must satisfy the requirement of statistical closeness to the data ``REAL''.
This combination of data can significantly increase the amount of data to be processed.
The method provides a unique opportunity to evaluate the parameters of the tumor, primarily the size and power of heat generated by the tumor.

\noindent 2) 
Using the combined dataset, tumors as small as 0.5 cm can be detected if they are in the rapid growth stage \cite{Gautherie-1982}, when volume doubling occurs in about 100 days or less.

\noindent 3) 
Convolutional neural networks for the ``SIMULATION'' sample give 71.5 percent accuracy in determining the location of the tumor based on the criterion of being in a given breast sector. We note the good agreement of this result with the estimates when using a multilayer perceptron network within 62--64 percent \cite{Glazunov-Polyakov-2021local}.

\noindent 4) 
An important feature of the MWR diagnostics is the ability to simultaneously have high values of sensitivity and specificity.
As a rule, mammography, ultrasound and MRI methods are better at detecting cancer patients (high sensitivity), but poorly at recognizing healthy people (low specificity). This is due to the difference in physical methods, when mammography, ultrasound and MRI are based on the determination of structural changes in tissues.
The MWR method detects temperature anomalies caused by inflammatory processes due to disease.

\noindent 5) 
We propose new 17-points breast examination scheme (instead of the traditional 9-point scheme), which allows you to build a better picture of temperature fields. Our analysis showed an increase in both sensitivity and specificity for this modified diagnostic algorithm.

Using only the ``REAL'' sample for the CNN does not give good results because the sample size is small and the number of Sick patients is also small, which is critical for CNN. The combined dataset provides satisfactory and close results for SVM and CNN (See Table~\ref{newmetodic}).
Note that the SVM algorithm turns out to be worse when working with feature spaces for breast oncology \cite{Polyakov-etal-2021ourVest}.

The 3D modeling makes it possible to create a models set with different internal structures of the breast. Each such model differs in the geometric parameters of such main components of the breast as the breast lobes, lactiferous sinus and ducts, adipose tissue, arterial and venous subsystems, and others. The size and shape of the breast also varies. The natural variability of the geometric characteristics of $\vec{\cal G}$ is quite large and the question of the influence of this uncertainty requires additional studies based on the construction of an even larger ``REAL'' sample, which apparently is necessary for processing breast cancer screening data when using MWR in practice. However, the wavelength for MWR is $\geq$ 2 cm, so the uncertainties in the sizes and localizations of biocomponents on smaller scales make a small contribution to the spatial temperature distributions. Thus, the statistical properties of ``REAL'' and ``SIMULATION'' datasets are consistent, providing a measurement error of about 0.2$^\circ$C \cite{Sedankin-Chupina-Vesnin-2018}.

The advantages and limitations of the proposed method for processing MWR thermometric data are due to both the method of microwave temperature measurement itself and the quality of dataset construction based on computer modeling. The ``SIMULATION'' sample quality is ensured by using realistic models of the internal structure of the breast, which contains all the main 3D components of this complex biotissue. Transition from the commonly used multilayer breast model \cite{Sedankin-etal-2018Mathematical, Zuluaga-Gomez-etal-2019thermography-breast-cancer, Ferreira-Numerical-analysis-2019, Mukhmetov-et-al-2021} to our multicomponent 3D internal structure seems to be a necessary step to build a better ``SIMULATION'' sample.
Multilayer models give formally higher values of sensitivity and specificity \cite{Mashekova-et-al-2022}, since the tumor as a hot zone stands out very well against the background of an almost uniform temperature distribution. Therefore, reliable detection of such a tumor is an artificial result due to the roughness of the approach, and such multilayer models poorly describe the real properties of the breasts.

We emphasize that the proposed method for applying combined datasets based on both real medical measurements and numerical simulations of MWR measurements can be effective for machine learning in solving a wide range of problems in diagnosing various organs and diseases.

\vspace{10pt}

\textbf{Funding:} This research was funded by the Ministry of Science and Higher Education of the Russian Federation (the government task no. 0633-2020-0003).
\vspace{6pt} 

\textbf{Acknowledgments:} The authors thank Ph.D. Tatiana Zamechnik (docent, Volgograd State Medical University) for providing expert advice when building  3D model of the internal structure of the breasts and prof. Alexander Losev for providing the dataset of real MWR examinations of breasts.
The authors thank anonymous referees for helpful comments, which improved the content of the paper.
\vspace{6pt}

\textbf{Conflicts of Interest:} The authors declare no conflict of interest. The funders had no role in the design of the study; in the collection, analyses, or interpretation of data; in the writing of the manuscript, or in the decision to publish the results.

\end{document}